\documentclass{aa}
\usepackage{graphicx}

\usepackage{natbib}
\bibpunct{(}{)}{;}{a}{}{,} 

\usepackage{array}
\usepackage{longtable}
\usepackage{textcomp}
\usepackage{multirow}
\usepackage{amstext}
\usepackage{esint}
\usepackage{url}
\usepackage{graphicx}
\usepackage{wrapfig}
\usepackage{lscape}
\usepackage{rotating}
\usepackage{epstopdf}
\usepackage{color, colortbl}
\usepackage{caption}

\makeatletter

\usepackage{multirow}
\titlerunning{An analysis of quasi-parallel and quasi-perpendicular acceleration}
\authorrunning{West et al.}

\makeatother

\usepackage{babel}

\begin{document}

\title{The connection between supernova remnants and the Galactic magnetic
field: An analysis of quasi-parallel and quasi-perpendicular cosmic ray acceleration
for the axisymmetric sample}

\author{J. L. West \inst{1} \and S. Safi-Harb
\thanks{Canada Research Chair}  
\inst{1}  \and G. Ferrand \inst{1} }
\institute{Dept of Physics and Astronomy, University of Manitoba, Winnipeg R3T 2N2, Canada\\
\email{jennifer.west@umanitoba.ca; samar.safi-harb@umanitoba.ca}
}

\date{Received 5 January 2016; accepted 8 September 2016}

\abstract {The mechanism for acceleration of cosmic rays in supernova remnants (SNRs) is an outstanding question in the field. We model a sample of 32 axisymmetric SNRs using the quasi-perpendicular and quasi-parallel cosmic-ray-electron (CRE) acceleration cases. The axisymmetric sample is defined to include SNRs with a double-sided, bilateral morphology, and also those with a one-sided morphology where one limb is much brighter than the other. Using a coordinate transformation technique, we insert a bubble-like model SNR into a model of the Galactic magnetic field. Since radio emission of SNRs is dominated by synchrotron emission and since this emission depends on the magnetic field and CRE distribution, we are able to simulate the SNRs emission and compare this to data. We find that the quasi-perpendicular CRE acceleration case is much more consistent with the data than the quasi-parallel CRE acceleration case, with G327.6+14.6 (SN1006) being a notable exception. We propose that SN1006 may be a case where both quasi-parallel and quasi-perpendicular acceleration are simultaneously at play in a single SNR.}


\maketitle

\section{\label{sec:intro}Introduction}

The mechanism for acceleration of cosmic rays in supernova remnants (SNRs) is an outstanding question in the field. A popular idea is that the distribution of the cosmic ray electrons (CREs) is responsible for determining the morphology of the so-called, bilateral
SNRs (e.g., \citealt{Petruk:2009bg,2011A&A...531A.129B,Reynoso:2013tr}). These are shell-type SNRs with two lobes of emission, separated by a symmetry axis. It has been long observed that many SNRs exhibit this type of morphology \citep[e.g.,][]{Kesteven:1987vt}.

There are primarily two acceleration scenarios that are considered in the literature:
quasi-perpendicular, where CREs are most efficiently accelerated when
the shock normal is perpendicular to the post-shock magnetic field;
and quasi-parallel where CREs are most efficiently accelerated when
the shock normal is parallel to the post-shock magnetic field (\citealp{Jokipii:1982jy,1989ApJ...338..963L,Fulbright:1990gu}
and references therein). The morphology of these two cases differs in that the axis of bilateral symmetry of the radio synchrotron emission is rotated by $90^\circ$ with respect to the direction of the ambient magnetic field. In the quasi-perpendicular case, the axis of bilateral symmetry is aligned with the ambient field, whereas in the quasi-parallel case the axis of bilateral symmetry is perpendicular to the ambient field as illustrated in Figure \ref{fig:CRE-geometry}. 

One of the main conclusions of \citet{Fulbright:1990gu} was that models of the quasi-parallel scenario produces images that are unlike any observed SNRs and thus, this study was supportive of the quasi-perpendicular case. Several studies since then have looked at these two cases, but there is disagreement  about which case is favored. For example, \citet{2011MNRAS.413.1657P} pointed out that the odd morphologies predicted by \citet{Fulbright:1990gu} would be expected to be fainter and less likely to be observed. Most of these recent studies have focused on G327.6+14.6 (SN1006), which is one of the brightest SNRs with excellent quality multi-wavelength data across the electromagnetic spectrum. This well studied, historical-type SNR has one of the most clearly defined bilateral structure of all SNRs. 

Two such studies present evidence in favor of the quasi-perpendicular case: \citet{Petruk:2009bg}, who compare the azimuthal brightness profile of radio maps to models, concluding that quasi-perpendicular injection is favored, and \citet{Schneiter:2010dh}, who compare both radio and X-ray emission to MHD models to conclude that the Galactic magnetic field (GMF) is most likely perpendicular to the Galactic plane.

Conversely, several other studies are supportive of the quasi-parallel scenario. These include \citet{Rothenflug:2004ch}, who suggest that the quasi-parallel scenario is a better fit on the basis of a geometrical argument regarding the limb-to-centre brightness ratios; \citet{2011A&A...531A.129B}, who compare the radio morphology to 3D MHD simulations and conclude that the bright limbs are polar caps; and most recently \citet{Schneiter:2015fc} who focus on a comparison of Stokes Q data to models of this polarization parameter. The addition of this extra observable, lead these authors to 
support the quasi-parallel case.

Simulations of diffusive shock acceleration (e.g., \citealt{2012ApJ...755..109M, 2013PhRvL.111u5003M, 2014ApJ...783...91C}) show that electrons are more efficiently accelerated in quasi-perpendicular shocks, whereas protons and other nuclei are more efficiently accelerated in
quasi-parallel shocks. Thus, it is proposed that one would expect to find large, turbulent magnetic fields downstream of parallel shocks, 
and simple, compressed fields in the case of quasi-perpendicular shocks. In terms of SN1006, these authors argue that the observations
of  \citet{Reynoso:2013tr} agree with their findings and favor the quasi-parallel, polar-cap, scenario of SN1006. However, they
also point out that further, detailed, multi-wavelength observations are necessary to conclusively prove this scenario.

As pointed out above, most studies on this topic in the context of SNRs have focussed
on SN1006 and we find no studies that undertake a global study of
these two CRE acceleration scenarios in the context of many other
SNRs with a similar, bilateral, appearance.

In this paper, we extend the work of \citet[][hereafter Paper 1]{2015arXiv151008536W}, which presents a detailed study of the radio morphology of 33 clearly-defined
SNRs with axisymmetric appearance\footnote{that includes double-sided bilateral shells as well as one-sided shells
where one limb is much brighter than the other}, and addresses whether this morphology can be reproduced by modeling these
SNR's appearance as being the result of compression of the Galactic magnetic field (GMF). Assuming an isotropic CRE distribution, Paper 1 finds a remarkable agreement (about 75\%) between SNR morphology and the model appearance when using the model of \citet[hereafter JF12]{Jansson:2012ep}, which is a model that includes an X-shaped vertical component. The agreement is much less convincing when an earlier GMF model \citep{Sun:2008bw}, one that does not include any vertical component, is tested.

Radiation from SNRs at radio wavelengths should be dominated by synchrotron radiation, the intensity of which is dependent on the CRE distribution as well as the magnetic field component that is in the plane of the sky. By using an isotropic CRE distribution, Paper 1 demonstrates that the morphology of clearly defined, axisymmetric SNRs is dominated by the large scale GMF. However, since an isotropic CRE distribution is not considered to be physically motivated and
the quasi-perpendicular and quasi-parallel CRE acceleration scenarios discussed above are usually considered more realistic \citep[e.g.,][]{Fulbright:1990gu}, it is important to extend the work of Paper 1, and consider the impact that the quasi-perpendicular and quasi-parallel CRE acceleration scenarios have on morphology for nearly the same sample of axisymmetric SNRs (with the exclusion of G001.9+00.3, discussed in the following section).

In Section~\ref{sec:Model} we summarize the modeling, which uses the same procedure as Paper 1. In Section~\ref{sec:Results} and Appendix~\ref{sub:Data-models} we present the results and discussion of this modeling. In section \ref{sec:SN1006}, we present a case study of SN1006 including further modeling (Section~\ref{sub:SN1006modeling}) and discussion (Section~\ref{sub:SN1006discussion}). G001.9+00.3, which is a special case similar to SN1006, is discussed in Section \ref{sec:g1-9}. Conclusions are found in Section~\ref{sec:conclusions}.

\section{\label{sec:Model}Model}

Paper 1 uses a coordinate transformation technique to insert a
bubble-like model SNR into a model of the GMF. Given the assumptions that the SNR is in the Sedov phase and that the magnetic field is frozen into the ambient plasma, this method appropriately drags the magnetic field lines of the GMF into the post-shock configuration. The Hammurabi Code%
\footnote{http://sourceforge.net/projects/hammurabicode/%
} \citep{Waelkens:2009bn} is then used to produce simulated Stokes I, Q, and U radio images, as well as polarized intensity ($PI = \sqrt {{Q^2} + {U^2}}$) and polarization angle ($PA=\frac{1}{2}\tan^{ - 1}\frac{U}{Q}$), which are then compared to real data. 

Images of all known Galactic SNRs are studied to choose the cleanest examples of SNRs with axisymmetric appearance.
Of these, 33 SNRs that are distributed around the Galaxy, are chosen and these are modeled at each of 11 
different distances: 0.5, 1, 2, 3, 4, 5, 6, 7, 8, 9, and 10~kpc from the Sun. These images are available on the companion website: Supernova remnant Models \& Images at Radio Frequencies (SMIRF: \url{http://www.physics.umanitoba.ca/snr/smirf/}). For this work, we divide this sample into two parts: those SNRs with a double-limbed, bilateral morphology and those with a single-limb. We make this separation since it can be argued that the single-limb SNRs may be interpreted as a double limb that has merged, which would impact the interpretation of the axial orientation. Thus we divide the sample into two parts so we can consider the single- and double-sided objects separately.

We have also looked at the ages of the SNRs in our sample. With the exception of G001.9+00.3 and possibly SN1006, all the SNRs (with known ages) are expected to be beyond the ejecta-dominated phase and well into the Sedov-Taylor phase of their evolution. Four of these are thought to be core-collapse (CC) type explosions (see Table \ref{tab:data-vs-model}) where the mass-loss of the progenitor may have perturbed the surrounding medium and magnetic field. However with the possible exception of G332.4--00.4, all of these SNRs are old enough and large enough that the compression of the ISM is expected to dominate. Two SNRs are thought to be Type Ia: G001.9+00.3, which is suspected to be Type Ia \citep{2008ApJ...680L..41R}, and SN1006, which is confirmed to be Type Ia (\citealp{1996ApJ...459..438S, 2003ApJ...585..324W}
and references therein). These two SNRs are also unique for other reasons (see below) and these are discussed separately in Section \ref{sec:SN1006}.

The focus is on {\em radio} data for the physical reason that we are dealing here with synchrotron emission from $\sim$GeV electrons. The inclusion of X-ray data complicates the interpretation since there is the possibility of confusion between the thermal X-ray emitting population and non-thermal emission from particles having TeV energies. We stress that our modeling here
is on the global morphology and applied to a selected sample of clean bilateral shells based on radio data, 
and so does not attempt to model small-scale structures, including knots of emission that would be associated, e.g., with instabilities in the shock and ejecta clumps \citep[e.g., ][]{2001ApJ...549.1119W}.

In addition, due to the availability of large scale radio surveys, there is more complete sample of SNRs that are relatively consistent in quality. In the case of X-ray data, in many cases there is no data available (or the data has very poor sensitivity) and where high quality data is available, it often only covers a portion of the SNR. The only cases where the X-ray data should be taken into consideration are those where it can be shown to be non-thermal. There are only four such cases in our sample that have non-thermal X-ray emission: G001.9+00.3, G028.6--00.1, G156.2+05.7, and SN1006 \citep[and references therein]{Ferrand:2012cr}. 

In the cases of G028.6--00.1 and G156.2+05.7 the X-ray emission is diffuse and does not exhibit the same bilateral morphology that is observed in radio. On the other hand, G001.9+00.3 and SN1006 are both very young and their X-ray emission does have a clear bilateral morphology. Thus these SNRs deserve special consideration and both are discussed later on in Section \ref{sec:SN1006}. G001.9+00.3 is the youngest in the sample and has the peculiar property that the brightest areas of X-ray and radio emission are anti-correlated. For this reason, we have chosen to exclude it from the sample under consideration, thus reducing our total sample size to 32 SNRs.

We use the same method and parameters as in Paper 1 and we refer the reader to that study for these additional details. As in Paper 1, we also present results for two GMF models those of JF12 and \citet{Sun:2008bw}.

The previous study assumes the simpler case of an isotropic CRE distribution.  For this study we include local acceleration effects, comparing the quasi-perpendicular and quasi-parallel CRE distributions. The most common way to model these two scenarios is to scale the CRE distribution by a factor that depends
on the angle between the shock normal and the post-shock magnetic
field, $\phi_{Bn2}$ (see \citealp{1989ApJ...338..963L,Fulbright:1990gu}).
In the quasi-parallel case this factor is given by $\cos^{2}\phi_{Bn2}$
and for the quasi-perpendicular case it is $\sin^{2}\phi_{Bn2}$. 

\begin{figure*}
\centering
\includegraphics[width=12cm]{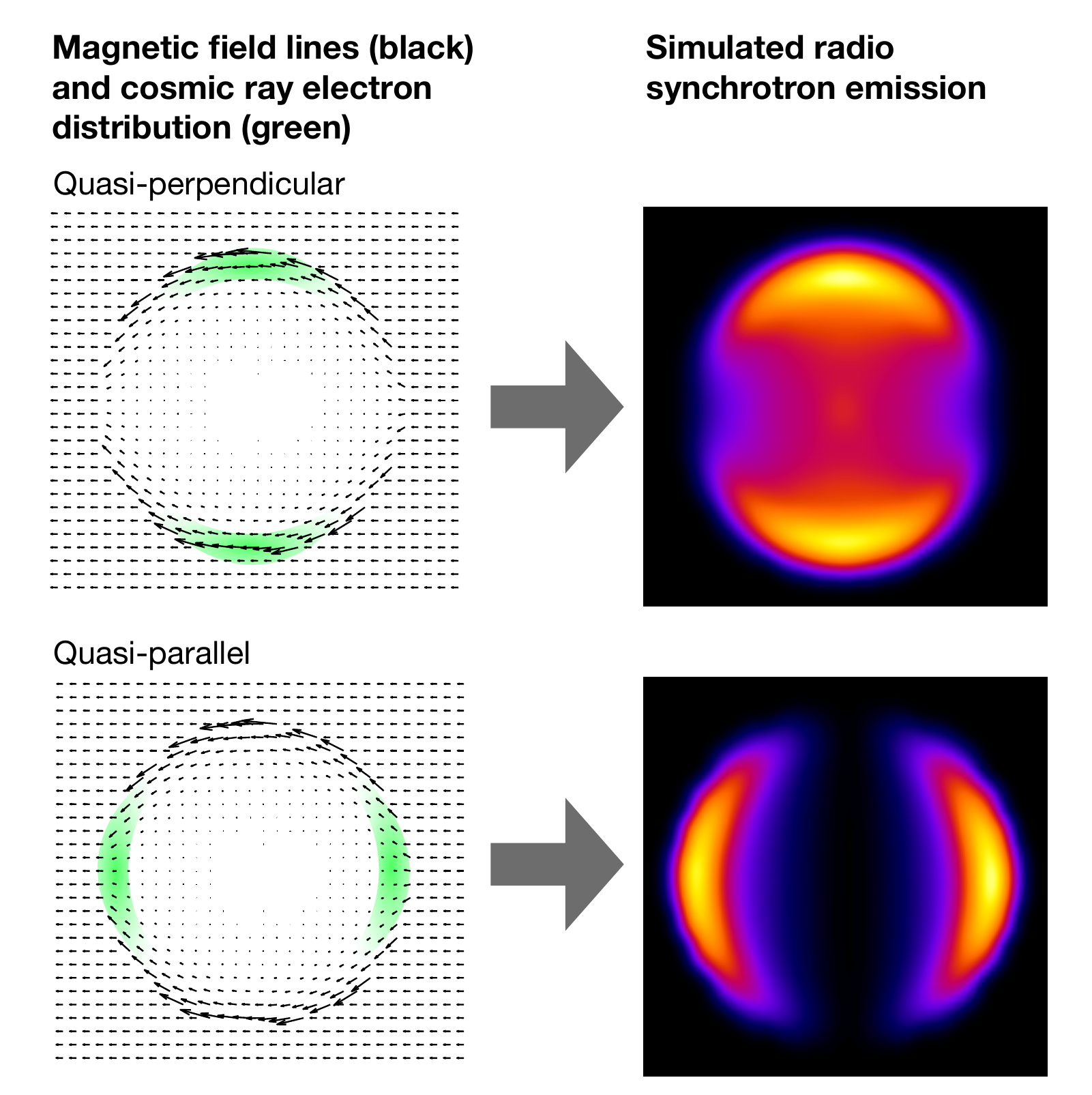}
\protect\caption{\label{fig:CRE-geometry}Geometry of CRE distributions for quasi-perpendicular shocks (top) and quasi-parallel shocks (bottom),
 and the corresponding
simulated synchrotron emission, which has been normalized for display purposes. This cartoon is intended to qualitatively show the distribution of the CREs with respect to the magnetic field geometry. It is not intended to be representative of the precise quantitative distributions. }
\end{figure*}

The quasi-perpendicular case is sometimes called the ``equatorial belt'', where the CREs are distributed around the region
where the magnetic field is subject to maximum compression. The quasi-parallel case is sometimes referred as ``polar caps'', since the CREs are distributed near the poles of a compressed magnetic field as illustrated in Fig.~\ref{fig:CRE-geometry}.


\section{\label{sec:Results}Results and Discussion}

Paper 1 describes in detail the process for selecting the sample of SNRs with axisymmetric appearance, which is briefly summarized here. The literature and data archives were searched to collect the best-available radio images of all known SNRs. From those, the cleanest and clearest examples of those with bilateral symmetry were selected. \citet{2007A&A...470..927O} showed that asymmetries in bilateral SNRs can be explained by gradients of ambient density or magnetic field strength and thus SNRs with a single well-defined limb are also included in the sample, however as mentioned above, in this paper we separate the single- and double-limbed SNRs into separate categories so they can be considered separately.

The results of the modeling are presented in Appendix \ref{sub:Data-models}. Here we show images  of the SNRs in comparison to the models for two CRE distributions, at all distances, and for the GMF of  JF12. This Appendix, while quite similar to Appendix D in Paper 1, has a very important difference since that figure showed models strictly for the isotropic case. By comparing to Paper 1, one can see that in terms of the morphology, the quasi-perpendicular case is very similar to the isotropic case for most models, although some small differences do exist, these are in terms of intensity differences and not overall morphology. More importantly, Appendix \ref{sub:Data-models} here provides a side-by-side comparison for the quasi-perpendicular and quasi-parallel CRE acceleration cases, which is important to visualize the significant morphological differences between these two cases.

Paper 1 uses a quantitative analysis to compare the models with the data using the bilateral axis angle, $\psi$, which is the angle between the axis of bilateral symmetry and the Galactic plane. Through the careful selection of the sample, the objective is to use only the clearest cases where the axis of bilateral symmetry can be unambiguously identified. For the double-limbed cases, the axis of bilateral symmetry is defined as the line running between the two limbs. The limbs are defined to be the brightest region of {\em radio} emission, {\em where a corresponding limb on the other side can be identified}. In some cases only a single limb is visible, where no counterpart can be detected on the opposite side. Here, the "symmetry" axis is chosen to be parallel to the brightest area of emission. This makes the assumption that there is a corresponding limb on the opposite side that is not detected. The angle is defined using a by-eye method due to the fact that the data is messy, with point sources and extraneous emission and it is therefore difficult to devise an automated method to measure this angle. In this study, $\psi$ is also difficult to define for many of the quasi-parallel models since several of them show a filled-centre morphology and thus, the angle of bilateral symmetry cannot be defined (e.g., G016.2--02.7, $d=2$~kpc). Therefore, we present the models and images together with a qualitative discussion.

The models show that for the quasi-perpendicular case, there is a reasonable morphological match between model and data for all cases examined and for some distance from 0.5 to 10~kpc, based on a qualitative comparison. The results for the quasi-parallel case, however, show that there are fewer matches between the models and the observations as the quasi-parallel case produces some strange morphologies that are not matched by any SNR, as was pointed out by \citet{Fulbright:1990gu}. In at least 10 out of the 32 cases, we find no reasonable morphological match between the data and the model. We also tested the quasi-parallel case for the GMF model of \citet{Sun:2008bw} and found an even poorer correspondence between the model and data than when compared to the JF12 model shown, which is consistent with the results of Paper 1.

The quasi-perpendicular case also has better consistency with published distances. While there are seven cases (G028.6--00.1, G093.3+06.9, G116.9+00.2, G119.5+10.2, G127.1+00.5, G327.6+14.6, and G332.4--00.4) where the quasi-parallel case has a reasonable morphological match, the distances in these cases are inconsistent with the published results. For these cases, the quasi-perpendicular case matches both in morphology and distance. There are three cases (G065.1+00.6, G156.2+05.7, G332.0+00.2) where the quasi-parallel case is consistent for both morphology and distance, but in the case of G156.2+05.7, the quasi-perpendicular case is also a reasonable match. G065.1+00.6 and G332.0+00.2 are the only two cases where the quasi-parallel case is consistent for both morphology and distance but the quasi-perpendicular case is not. In the case of G065.1+00.6, the best fit quasi-perpendicular case is not in agreement with the published distance of 9.0--9.6 kpc \citep{Tian:2006bp}, however, this published distance was based on a possible association with HI emission that has yet to be confirmed. See Table \ref{tab:data-vs-model} for a summary of the results of the distances.

\begin{figure*}
\centering
\includegraphics[width=17cm]{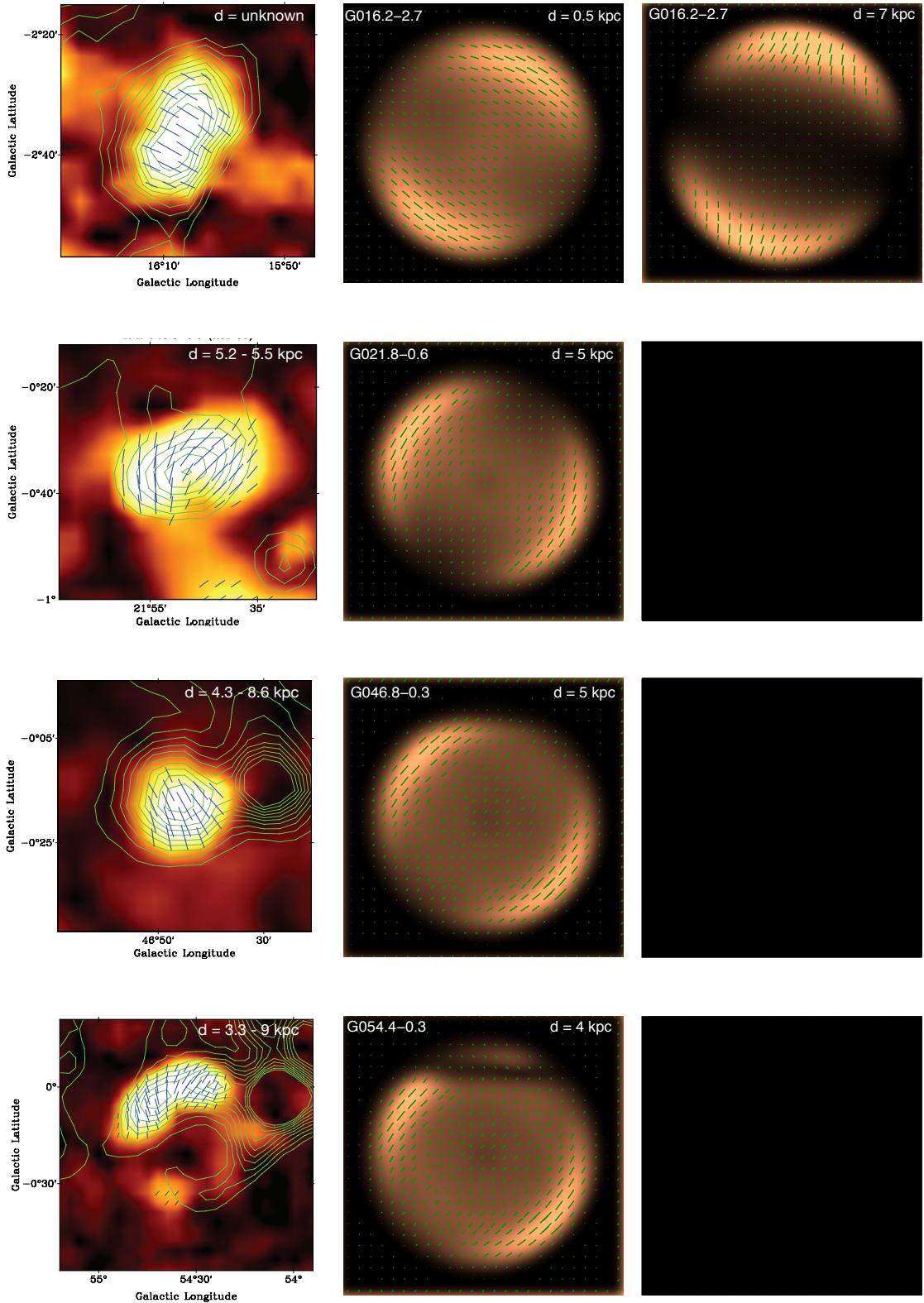}
\protect\caption{\label{fig:B-fields}Comparison of the magnetic fields for all cases where a magnetic field has been measured. Magnetic field vectors are plotted on top of polarized intensity emission. Data (left column) from top to bottom: G016.2--02.7 \citep{Sun:2011jz}, G021.8--00.6 \citep{Sun:2011jz}, G046.8--00.3 \citep{Sun:2011jz}, G054.4--00.3 \citep{Sun:2011jz}. Where the data is presented in equatorial coordinates, it has been rotated to Galactic coordinate for consistency with the models. Centre: Best-fit quasi-perpendicular case using the JF12 model. Right: Best-fit quasi-parallel case using the JF12 model. In some cases there are no reasonably matching models and the model is shown as blank in these cases.}
\end{figure*}

We searched the literature to find all available magnetic field observations of this sample of 32 SNRs to compare to the magnetic field predicted by our best fitting models. In Figure \ref{fig:B-fields}, we present the data and the models. In 13 out of 15 cases, the observed magnetic field is more consistent with the quasi-perpendicular model. Moreover, in all but one of these cases, the distance is also in agreement with the best  quasi-perpendicular model within our uncertainty. G065.1+00.6 is again the one case that disagrees in terms of distance, but as discussed above, the published distance to G065.1+00.6 is unconfirmed.

The other cases where the direction of the magnetic field is inconsistent are G046.8--0.3 and G327.6+14.6 (SN1006). In the case of G046.8--0.3, polarization data reveals a radial magnetic field (see Figure \ref{fig:B-fields}), however this data is very low resolution compared to the best available radio image (see image in Figure \ref{fig:data-models-double}). As noted in Paper 1, the simulated polarization vector plot using the JF12 model is tangential, but it is interesting that the plot using the Sun et al. (2008) model at the corresponding distance (4 kpc) does show a radial magnetic field (see Figure 7 of Paper 1). The case of SN1006 is discussed in the following section. It should also be noted that SN1006 is the only case where the magnetic field predicted by the quasi-parallel model is at all consistent with the observations.

Overall, these results are consistent with the morphology of clean, axisymmetric SNRs being the result of quasi-perpendicular shocks in a simple, compressed GMF. 

\subsection{\label{sec:Brightness}Brightness parameters}

Other authors \citep[e.g.,][]{Petruk:2009bg, Rothenflug:2004ch} have used brightness parameters such as the limb-to-centre ratio and the radial brightness profiles as a variable to help distinguish between the two CRE scenarios, and we looked at the feasibility of using these parameters in the context of this study.

\citet{Rothenflug:2004ch} use a geometrical argument to claim that the quasi-parallel scenario is the only one consistent with observations. They argue that in the quasi-perpendicular scenario, the limb-to-centre brightness ratio must be at least 0.5 because emission is coming from the equatorial belt around the whole perimeter of the SNR. Since the X-ray observations of SN1006 reveal a ratio that is smaller than this (i.e., 0.3) they claim that the quasi-perpendicular scenario is ruled out. However, this study also showed that the radio data of SN1006 has a ratio of 0.7, which is quite different from the X-ray observations. Based on the radio data alone, this ratio is not inconsistent with the quasi-perpendicular scenario. Since this study is focussed on radio observations, and non-thermal X-rays are not observed in nearly all the SNRs in the sample (see previous section), the limb-to-centre brightness ratio is not useful for distinguishing between the two scenarios in the cases we are studying.

One other issue with this argument in the context of this study, is that the \citet{Rothenflug:2004ch} argument is only valid for the case of isotropic synchrotron emission found in a region with a disordered magnetic field. In this study, we are assuming an ordered field, and in addition, its initial orientation could include changes of direction (such as a bend) within the region where the SNR is inserted. 
For the data, there are also other associated uncertainties in determining the value of the limb-to-centre ratio. This variable could be affected by a non-uniform background level and the presence of regions of the ISM of varying density where emission can be enhanced. The localization of these regions is difficult to determine, but it is reasonable to assume that they will not be uniform around the equatorial belt and these enhancements could occur anywhere along the line-of-sight, in front or behind the SNR. Additionally, all of the radio data in this study is interferometry data, which in many cases is lacking the addition of short spacings information and thus impacting the limb-to-centre ratio. These factors combined with the uncertainty introduced by the magnetic field's directional dependence on the synchrotron emission means that this parameter is not useful for this study.

\citet{Petruk:2009bg}, use azimuthal brightness profiles to favour the quasi-perpendicular scenario, however these profiles can depend on the specific CRE distribution model used. In addition, this model does not account for magnetic field amplification, which is known to modify the non-thermal emission. For these reasons, the azimuthal brightness is also not useful for discriminating between the two CRE distribution scenarios.

Instead of using these quantitative ratios, this study takes the approach of a qualitative analysis that includes the orientation and magnetic field information.



\section{\label{sec:SN1006}Case study: SN1006}
Being a very bright, historical-type SNR with a very well-defined bilateral structure, SN1006 has been the subject of many previous studies, and thus it is important to address this SNR in particular. SN1006 has an observed a diameter of  $30'$ and a distance of 1.6--2.2~kpc \citep[and references therein]{Ferrand:2012cr}, although \citet{2013Sci...340...45N} set an upper limit on the distance of 2.1~kpc. The remnant is oriented at an angle of $83^\circ  \pm 5^\circ$ with respect to the
Galactic plane, which has led to controversy over whether the ambient magnetic
field local to SN1006 is oriented perpendicular or parallel to the Galactic
plane, depending on whether the quasi-perpendicular or quasi-parallel case is
favored by the particular study.

Our models presented in Appendix~\ref{sub:Data-models} show that for the
quasi-perpendicular case we find a very reasonable morphological fit at a distance of 1$\pm$1~kpc, which is in agreement with the range of published distances and given the uncertainties in the JF12 model. However, for the
quasi-parallel case, there is no reasonable fit for any distance modeled,
although the 0.5~kpc case is the closest match in terms of orientation.


\subsection{\label{sub:SN1006modeling}Further modeling of SN1006}

Any model of SN1006 must be able to account for all observations, including observations of radio polarization. \citet{Reynoso:2013tr} published detailed observations of this SNR that show that the magnetic field is radial in appearance. The availability of the Stokes Q radio polarization parameter from Reynoso et al.'s (2013) observations led \citet{Schneiter:2015fc} to model this additional parameter and conclude that the quasi-parallel case was closer to observations than the model for the quasi-perpendicular case. This conclusion, based on these new observations, was contrary to earlier results from the same authors \citep{Schneiter:2010dh} that supported the quasi-perpendicular case, highlighting the need to include all available observations. 

The Hammurabi code also provides model Stokes Q and U images, which together may be used
to produce a simulation of the magnetic field vectors that would be observed. We
compare our models simultaneously to the following observables: total intensity
(Stokes I) emission, the polarized intensity emission, and the magnetic field
vectors (i.e., polarization angle $+90^\circ$). 

The JF12 model likely does a good job of modeling the global properties of the
GMF, however local variations are not included so it is possible that the model will not be correct at the position of some individual SNRs. If we assume this
to be the case for the position of SN1006, then the magnetic field may have some
arbitrary configuration. We model SN1006 using a uniform magnetic field that is
defined by $B_x$ (line-of-sight component), $B_y$ (horizontal component), and
$B_z$ (vertical component) to find a model that takes into account all of the
observable properties described above.

We note that since SN1006 is tilted by $83^\circ$ with respect to the Galactic plane, the $B_y$ and $B_z$ components must have a specific relationship in order to give the model SNR the same angle. For the quasi-perpendicular case, $\tan(\psi) = B_z/B_y$ and for the quasi-parallel case, $\tan(\psi) = -B_y/B_z$. For example, in our models of the quasi-parallel case, $B_y$ is set to 1 $\mu$G, which means $B_z$ must be 0.12 $\mu$G to give the correct orientation whereas in the quasi-perpendicular case, $B_z$ is set to 1 $\mu$G, which means $B_y$ must be -0.12 $\mu$G for the same reason. The absolute values of $B_y$ and $B_z$ are not significant; it is the ratios that matters since we are interested in a more qualitative analysis.

By altering the $B_x$ component, one changes the centre-to-limb brightness ratio and the radial brightness profiles. In Figure \ref{fig:SN1006-models}, we show several quasi-parallel and quasi-perpendicular models that have the correct orientation, but with varying values of $B_x$.  

In the quasi-parallel case, $B_x$ must be less than $B_y$, otherwise the centre of SNR is completely filled in and the limbs are not defined. We show two cases that show that the pattern of the magnetic field vectors is radial in the limbs, as in the data, and remains essentially unchanged for all cases where $B_x<B_y$.

In the quasi-perpendicular case, the pattern of the magnetic field vectors are
tangential to the limbs for most cases, except when the $B_x$ component gets large (Figure \ref{fig:SN1006-models}, bottom row), the magnetic field pattern changes to a radial pattern. The morphology of the polarized intensity emission also changes significantly for this case, becoming inconsistent with the data. Also, the morphology of the emission changes from bright limbs to ring-like. In Figure \ref{fig:SN1006-models}, we show three  quasi-perpendicular cases showing this transition from tangential to radial magnetic field and illustrating that it is not possible to match both the intensity pattern and the magnetic field at the same time. 

We do not attempt a detailed quantitative optimization of parameters such as comparison of the limb-to-centre brightness ratio or radial brightness profiles. Such analyses are done in previous works (see discussion in Section~\ref{sec:intro}) with conflicting conclusions as to which CRE acceleration scenario is supported. Instead, we present our models here for qualitative comparison between the morphology together with the magnetic field pattern.

\subsection{\label{sub:SN1006discussion}Discussion of SN1006}

Our models agree with \citet{Schneiter:2015fc} in that the quasi-parallel model
most closely matches the complete set of observations. Thus, we conclude that the quasi-parallel case is
indeed the better fit in the case of SN1006, which is different from the results for most of the other
SNRs in this study. In the majority of the other cases the quasi-perpendicular case has better fit. 

We do note that the quasi-parallel case is not perfect in describing the observations. In particular, the
magnetic field vectors for the quasi-parallel model converge at the poles but
this is not observed in the high resolution magnetic field vector map by
\citet{Reynoso:2013tr}. 

It may be that SN1006 and possibly G001.9+00.3 as well (discussed below), both being very young SNRs may be different. Young SNRs would be expected to be dominated by turbulence and it is likely that in these young cases the Rayleigh-Taylor instabilities are playing a major role.

The magnetic fields of SN1006, and other
young SNRs, are seen to be amplified at the bright limbs
(\citealp{2003ApJ...589..827B,2004AdSpR..33..376B,2006AdSpR..37.1439B,
2012SSRv..166..231R}). Our models have not accounted for
this turbulent amplification, nor do those of other authors 
\citep[e.g.,][]{Schneiter:2015fc}. It has been suggested that  turbulence can lead to
selective amplification of the radial component of the magnetic field
\citep[e.g.,][]{Inoue:2013el}. We note that this could alter the observed properties,
particularly the polarized emission and observed magnetic-field vectors. This
turbulence must not be so significant as to destroy the regular, bilateral
appearance, but it may be possible that it imparts the apparent radial magnetic
field pattern.

Furthermore, observations of SN1006 show non-thermal X-ray and $\gamma$-ray emission (\citealp{1995Natur.378..255K,2010A&A...516A..62A}) in correspondence to the bright limbs. \citet{2014ApJ...783...91C} suggest that the $\gamma$-ray emission could indicate ion-acceleration and thus could further support the quasi-parallel scenario given the prediction that ions are more efficiently accelerated in quasi-parallel shocks.

We have the following observations: 

1. that the quasi-parallel case is favored for SN1006, 

2. that the quasi-perpendicular case is favored for the majority of other axisymmetric SNRs, 

3. that the JF12 model gives
a good fit to the total intensity emission for the quasi-perpendicular CRE case
and for a distance that is consistent with other measurements, 

4. that neither the quasi-parallel nor quasi-perpendicular cases can convincingly reproduce the observed pattern of the magnetic field, and

5. that in the case of SN1006, there is disagreement in the literature as to whether the quasi-parallel or quasi-perpendicular CRE case is favored with 
evidence supporting both scenarios.

We therefore suggest that the regular bilateral morphology of SN1006 may be due to the compressed GMF, and that this compression is around the
equatorial belt. The observed radial magnetic field pattern may be imparted by turbulence that does not destroy the morphology that is due to the direction of the regular component of the field. 

Additionally, we suggest that it is possible that both quasi-parallel and quasi-perpendicular  
acceleration could be simultaneously at play together in the same SNR. That is, the initial, compression of the GMF leads to quasi-perpendicular acceleration of electrons, which leads to the synchrotron radio emission. Then, turbulence in the young SNR may lead to local magnetic field amplification, that results in a radially oriented magnetic field component. At this point, quasi-parallel acceleration can occur, leading to the acceleration of ions and the the presence of the $\gamma$-ray emission at the limbs, coincident with the radio emission. 

\subsection{\label{sec:g1-9} Discussion of G001.9+00.3}

G001.9+00.3 is a very young SNR, thought to be the youngest in the Galaxy at only 150 -- 220 yr \citep{2011ApJ...737L..22C, 2009RMxAA..45...91G, 2008MNRAS.387L..54G, Reynolds:2008fy} and located at a distance of 8.5~kpc \citep{Reynolds:2008fy}.  This SNR has the peculiar property that the brightest areas of X-ray and radio emission are anti-correlated. Using the method described above (where the axis of bilateral symmetry is defined as being parallel to the brightest region of radio emission, where a corresponding limb on the other side can be identified) the axis of symmetry would be inclined at a large angle ($\sim-85^\circ$), making it nearly vertical. However, looking at the X-ray data \citep[see][]{Reynolds:2008fy}, which in the case of this SNR is mostly non-thermal, it is clear that the X-ray data shows a symmetry axis that is close to parallel with the Galactic plane. 

For completeness, we present our models compared to radio data in Figure \ref{fig:G1.9}. The extreme youth of this SNR means that is almost certainly has not reached the Sedov phase of its evolution, making it a unique case where this modeling does not apply.

\section{\label{sec:conclusions}Conclusions}
Our results support the conclusion of \citet{Fulbright:1990gu} that the
quasi-parallel scenario produces images that are unlike any observed
SNRs. We show that the large majority of our models, about 75\%, have a good match to the data for the quasi-perpendicular CRE case and most are consistent with published distance estimates to the SNRs. In the quasi-parallel CRE case, there is very poor agreement between the appearance of the models and the data. Thus we conclude that the radio morphology of most axisymmetric SNRs is the result of quasi-perpendicular shocks in a simple, compressed GMF.

SN1006 and G001.9+00.3 are notable exceptions, and in the case of SN1006 we find that neither of the simple, quasi-parallel nor quasi-perpendicular CRE cases, can fully describe the observations; although given a single choice, the quasi-parallel CRE case is the better option. Given that the JF12 GMF model with the quasi-perpendicular CRE case gives a good fit to the morphology of SN1006 at a distance that is consistent with published distance to SN1006, we suggest that the regular radio morphology of this SNR, like the majority of other axisymmetric SNRs, is due to the compressed GMF, and that this compression is around the equatorial belt. The deviations from the quasi-perpendicular model, which includes the observed radial magnetic field pattern, may be imparted by a turbulent magnetic field component that is in addition to the regular magnetic field component. We suggest that SN1006, and possibly other young, energetic SNRs that show non-thermal X-ray and $\gamma$-ray emission, are examples where both quasi-parallel and quasi-perpendicular acceleration could be simultaneously at play.

\begin{acknowledgements}
This research was primarily supported by the Natural Sciences and
Engineering Research Council of Canada (NSERC) through a Canada Graduate
Scholarship (J. West) and the Canada Research Chairs and Discovery
Grants Programs (S. Safi-Harb). The modeling was performed on a local
computing cluster funded by the Canada Foundation for Innovation (CFI) and
the Manitoba Research and Innovation Fund (MRIF). \\
\\

The radio images presented in this paper and companion website made
use of data obtained with the following facilities: CHIPASS 1.4 GHz
radio continuum map, Canadian Galactic Plane Survey (CGPS) and other
data from the Dominion Radio Astrophysical Observatory, Southern Galactic
Plane Survey, Effelsberg 100m Telescope and Stockert Galactic plane
survey (2720 MHz) (via MPIfR's Survey Sampler), Molonglo Observatory
Synthesis Telescope (MOST) Supernova Remnant Catalogue (843 MHz),
Sino-German 6 cm survey. Additionally we acknowledge the use of NASA's
SkyView facility (http://skyview.gsfc.nasa.gov) located at NASA Goddard
Space Flight Center for the data from the 4850 MHz Survey/GB6 Survey,
NRAO VLA Sky Survey, Sydney University Molonglo Sky Survey (843 MHz),
and the Westerbork Northern Sky Survey (325 MHz Continuum). Very Large
Array (VLA) data was acquired via the NRAO Science Data Archive and
the Multi-Array Galactic Plane Imaging Survey. We thank Michael Bietenholz
and David Kaplan for providing the VLA images of G021.6--00.8 and G042.8+00.6,
respectively. \\
\\
This research made use of APLpy, an open-source plotting package for Python (http://aplpy.github.com) and Astropy, a community-developed core Python package for Astronomy (Astropy Collaboration, 2013). This research also made use of Montage, which is funded by the National Science Foundation under Grant Number ACI-1440620, and was previously funded by the National Aeronautics and Space Administration's Earth Science Technology Office, Computation Technologies Project, under Cooperative Agreement Number NCC5-626 between NASA and the California Institute of Technology.\\
\\
This publication uses data generated via the Zooniverse.org platform, development of which was supported by a Global Impact Award from Google, and by the Alfred P. Sloan Foundation.\\
\\
This research has made use of the NASA Astrophysics Data System (ADS).

\end{acknowledgements}

\bibliographystyle{aa}
\bibliography{references}

\begin{figure*}
\centering
\includegraphics[width=17cm]{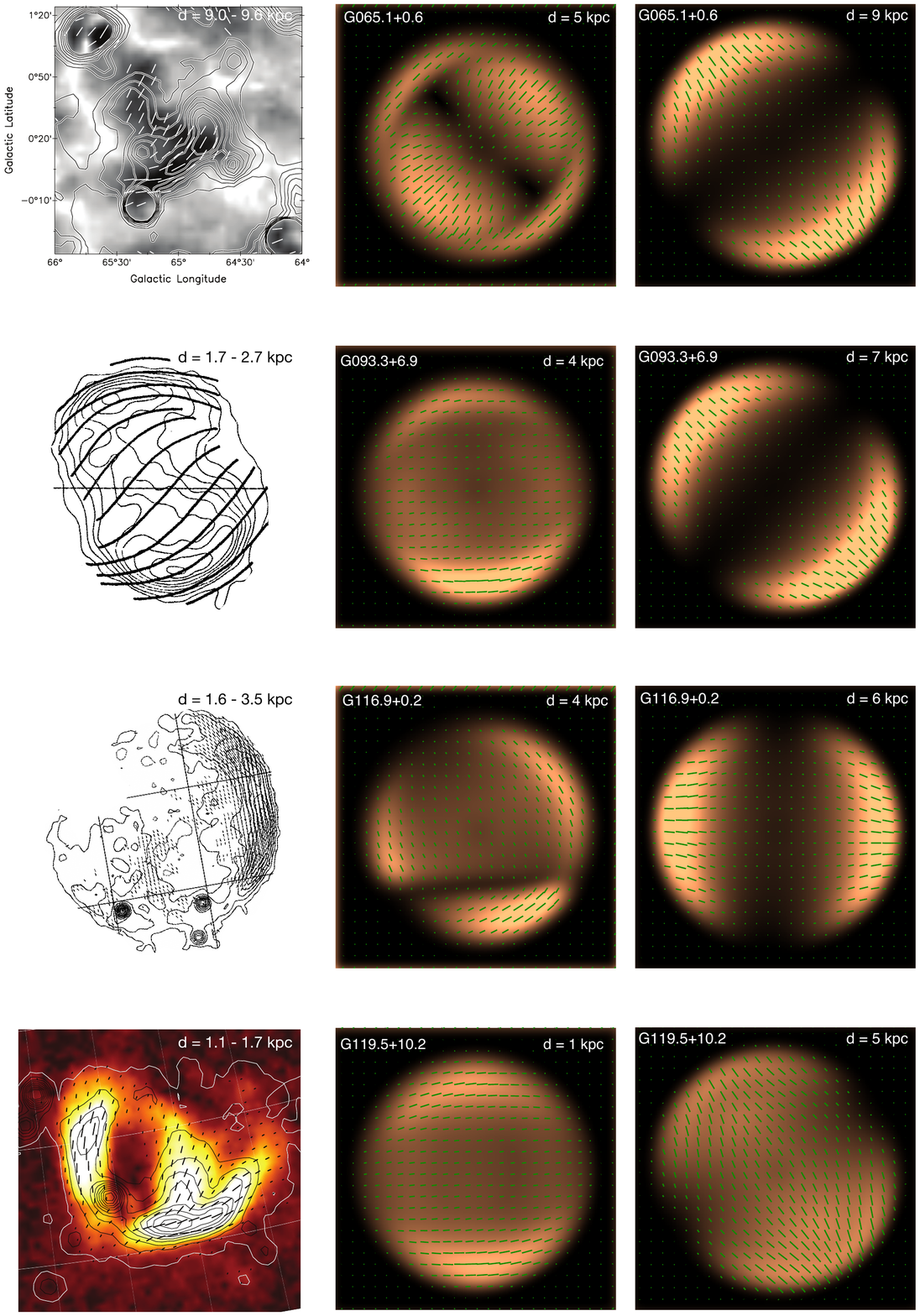}
\\
Fig. \ref{fig:B-fields} continued. Data (left column) from top to bottom: G065.1+00.6 \citep{Gao:2011fw}, G093.3+06.9 \citep{Milne:1987wi}, G116.9+00.2 \citep{2002nsps.conf....1R}, G119.5+10.2 \citep{Sun:2011hr}.
\end{figure*}

\begin{figure*}
\centering
\includegraphics[width=17cm]{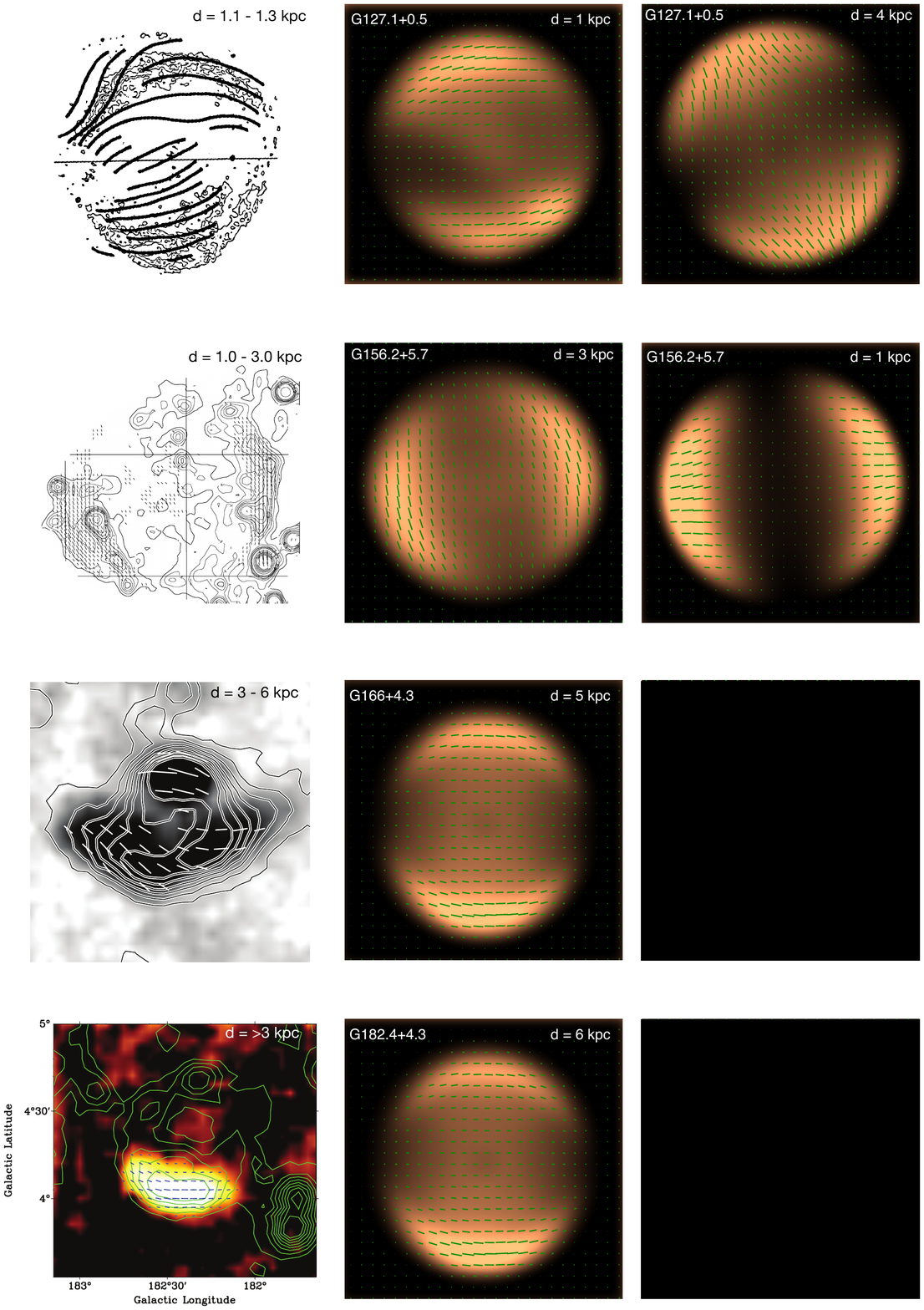}
\\
Fig. \ref{fig:B-fields} continued. Data (left column) from top to bottom: G127.1+00.5 \citep{Milne:1987wi}, G156.2+05.7 \citep{Reich:1992wc}, G166.0+04.3 \citep{Gao:2011fw}, G182.4+04.3 \citep{Sun:2011jz}.
\end{figure*}

\begin{figure*}
\centering
\includegraphics[width=17cm]{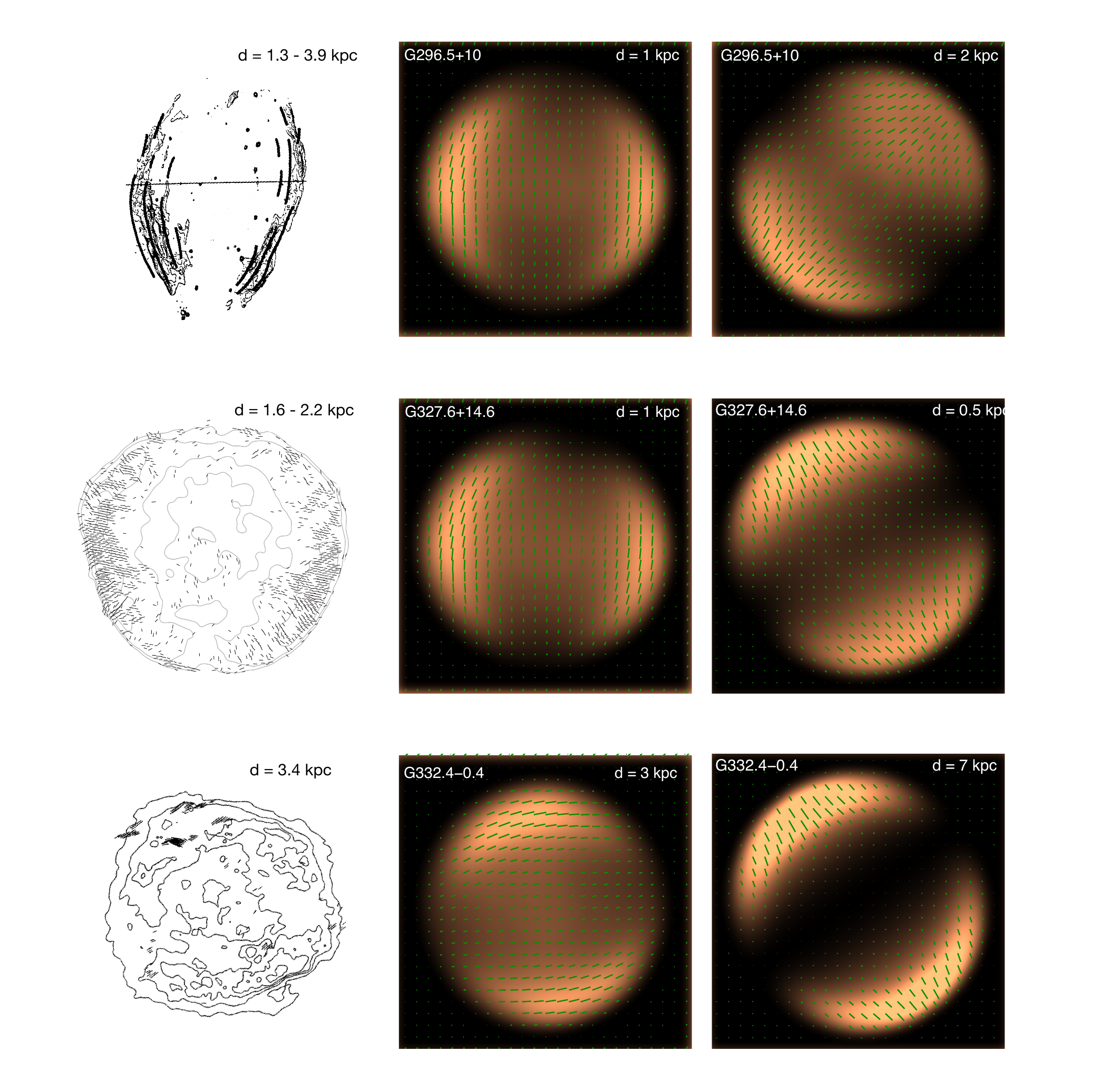}
\\
Fig. \ref{fig:B-fields} continued. Data (left column) from top to bottom: G296.5+10.0 \citep{Milne:1987wi}, G327.6+14.6 \citep{Reynoso:2013tr}, G332.4--00.4 \citep{1996AJ....111..340D}.
\end{figure*}

\begin{figure*}
\centering
\includegraphics[width=11cm]{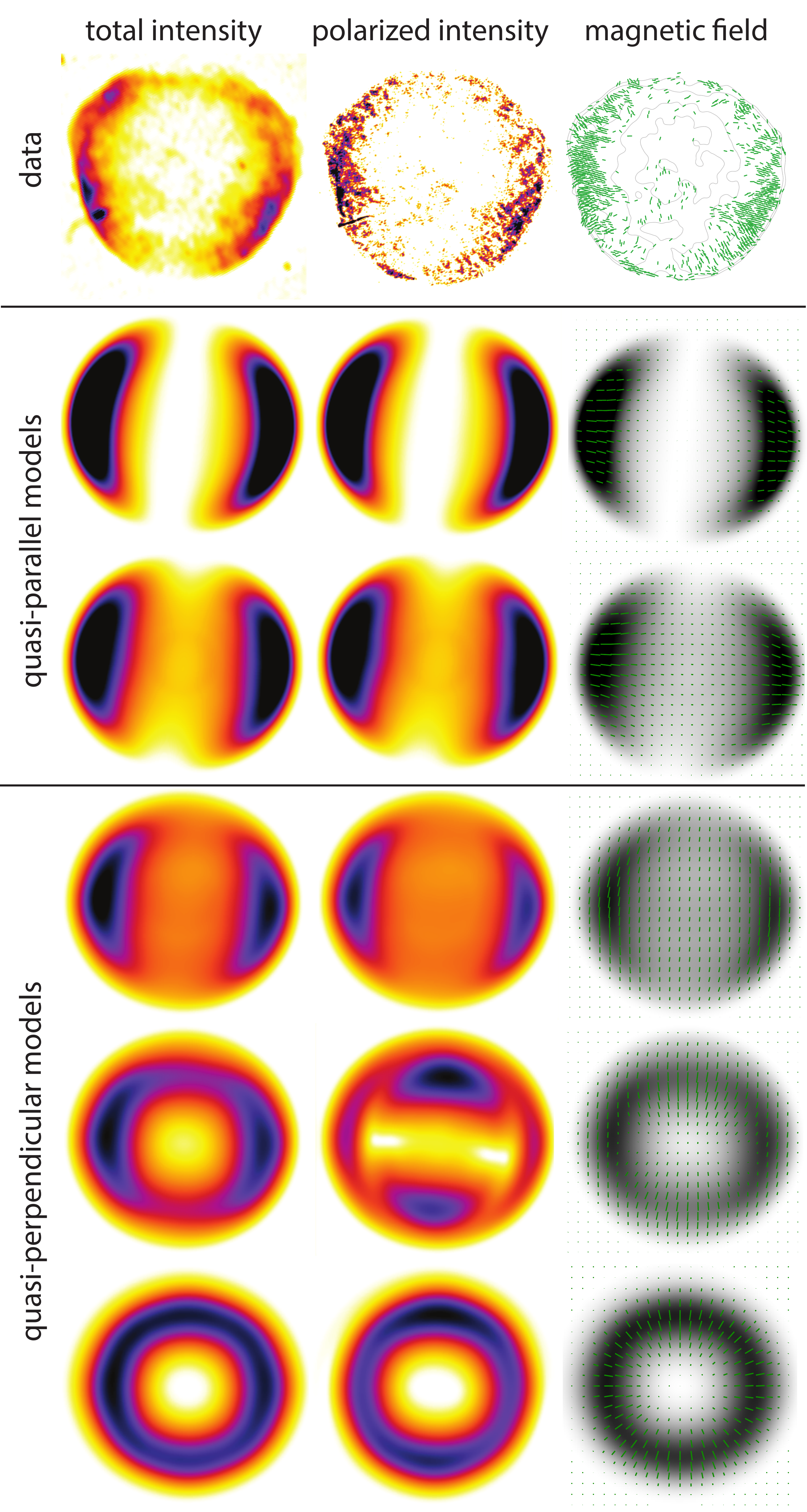}
\protect\caption{\label{fig:SN1006-models}Observations of G327.6+14.6 (SN1006) compared to a selection of grid-based models. The magnetic field is defined by the components $B_x$ (line-of-sight component), $B_y$ (horizontal component), and $B_z$ (vertical component), where it is the ratio between the $B_y$ and $B_z$-components that determines the orientation of the bilateral symmetry axis.
Top row: Observations of SN1006. Left: Stokes I total intensity (MOST). Centre: polarized intensity (Reynoso et al. 2013). Right: Magnetic field vectors shown with total intensity contours in the background (adapted from Reynoso et al. 2013). 
2nd row: Quasi-parallel model with $B_x = 0.03$ $\mu$G, $B_y = 1.0$ $\mu$G, and $B_z = 0.12$ $\mu$G.
3rd row: Quasi-parallel model with $B_x = 0.55$ $\mu$G, $B_y = 1.0$ $\mu$G, and $B_z = 0.12$ $\mu$G. 
4th row: Quasi-perpendicular model with  $B_x = 1.28$ $\mu$G, $B_y = -0.12$ $\mu$G, and $B_z = 1.0$ $\mu$G. This is very close to the case at the location of SN1006 in the JF12 GMF model.
5th row: Quasi-perpendicular model with  $B_x = -3.84$ $\mu$G, $B_y = -0.12$ $\mu$G, and $B_z = 1.0$ $\mu$G. 
Bottom row: Quasi-perpendicular model with  $B_x = -12.5$ $\mu$G, $B_y = -0.12$ $\mu$G, and $B_z = 1.0$ $\mu$G.
The models are arranged as the data with far left: Stokes I total intensity, centre: polarized intensity, and right: magnetic field vectors shown over total intensity emission. }
\end{figure*}


\begin{figure*}
\centering
\includegraphics[width=17cm]{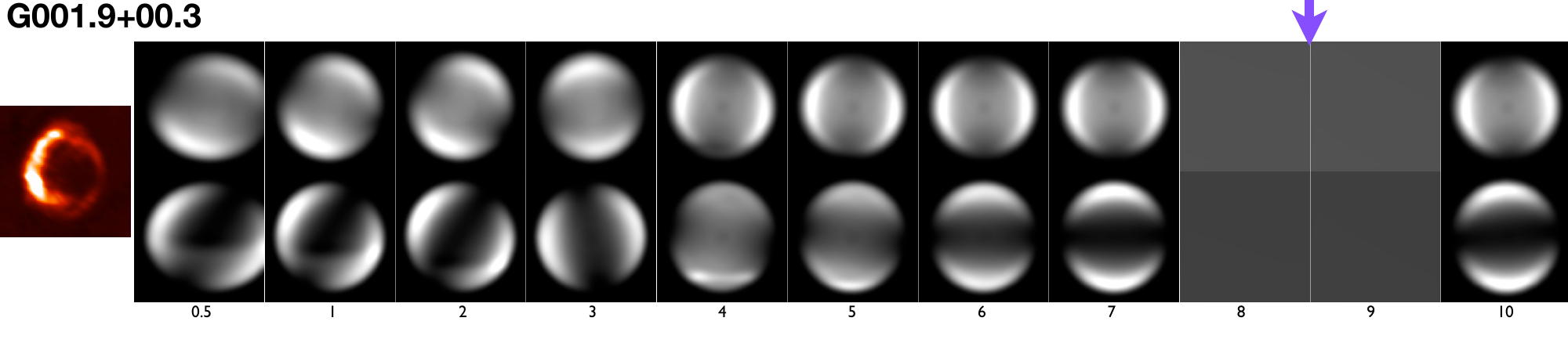}
\protect\caption{\label{fig:G1.9}Models of the SNR G001.9+00.3 compared to VLA radio data from \citet{Reynolds:2008fy}. The arrow marks the best published distance estimate of 8.5~kpc \citep[and references therein]{Ferrand:2012cr}}
\end{figure*}

\newpage

\onecolumn

\begin{table}
\caption{\label{tab:data-vs-model}Summary of the published parameters for the SNRs chosen for the study shown with distance ranges corresponding to models having a reasonable morphological match.}
\begin{tabular}{l|ll|ccc|c}
\multicolumn{3}{l}{
\multirow{2}{*}{SNR} } & \multicolumn{3}{c}{Distance {[}kpc{]}} & \multirow{2}{1.0cm}{Image Ref.}\tabularnewline
 & Age (yr)  & Type  & Published  & Q-Perp  & Q-Par  & \tabularnewline
\hline 
\hline 
\multicolumn{7}{|c}{Double-limbed}\tabularnewline
\hline 
\hline 
G003.7--00.2  & ?  &  & unknown  & $3_{-1}^{+1}$  & $1_{-2}^{+2}$  & 1\tabularnewline
G008.7--05.0  & ?  &  & unknown  & $2_{-2}^{+1}$  & $9_{-1}^{+2}$  & 2\tabularnewline
G016.2--02.7  & ?  &  & unknown  & $0.5_{-0.5}^{+2.5}$  & $7_{-3}^{+4}$  & 2\tabularnewline
G021.8--00.6  & 4400--5000  &  & 5.2--5.5  & $5_{-1}^{+6}$  &  & 3\tabularnewline
G028.6--00.1  & $\le$2700  &  & 6--8.5  & $6_{-2}^{+5}$  & $1_{-2}^{+3}$  & 3\tabularnewline
G036.6+02.6  & ?  &  & unknown  & $2_{-2}^{+2}$ {[}or $8_{-3}^{+3}${]}  &  & 2\tabularnewline
G046.8--00.3  & ?  &  & 4.3--8.6  & $5_{-2}^{+4}$  &  & 3\tabularnewline
G054.4--00.3  & 61000  &  & 3.3--3.5  & $4_{-2}^{+2}$  &  & 4\tabularnewline
G065.1+00.6  & 40000--140000  &  & 9.0--9.6  & $5_{-2}^{+2}$  & $9_{-1}^{+2}$  & 4\tabularnewline
G093.3+06.9  & 5000--7000  &  & 1.7--2.7  & $4_{-2}^{+2}$  & $7_{-2}^{+4}$  & 5\tabularnewline
G116.9+00.2  & 7500--15000  & CC  & 1.6--3.5  & $4{}_{-1}^{+2}$  & $6_{-1}^{+2}$  & 4\tabularnewline
G119.5+10.2  & 13000  & CC  & 1.1--1.7  & $1_{-1}^{+3}$  & $5_{-2}^{+5}$  & 6\tabularnewline
G127.1+00.5  & 20000--30000  &  & 1.1--1.3  & $1_{-1}^{+3}$  & $4_{-1}^{+1}$  & 4\tabularnewline
G156.2+05.7  & 7000--26000  &  & 1.0--3.0  & $4_{-2}^{+1}$ {[}or $8_{-2}^{+3}${]}  & $1_{-2}^{+2}$  & 7\tabularnewline
G296.5+10.0  & 3000--20000  & CC  & 1.3--3.9  & $1_{-0.5}^{+1}$  & $2_{-1}^{+1}$  & 8\tabularnewline
G302.3+00.7  & ?  &  & unknown  & $7_{-3}^{+3}$  &  & 8\tabularnewline
G317.3--00.2  & ?  &  & unknown  & $1_{-1}^{+2}$  &  & 8\tabularnewline
G321.9--00.3  & ?  &  & unknown  & $8_{-4}^{+3}$ {[}or $1_{-1}^{+2}${]}  &  & 8\tabularnewline
G327.6+14.6  & 1010  & Ia  & 1.6--2.2  & $1_{-0.5}^{+1}$  & $0.5_{-0.5}^{+0.5}$  & 8\tabularnewline
G332.0+00.2  & ?  &  & >6.6  & $1{}_{-1}^{+2}$  & $7_{-3}^{+4}$  & 8\tabularnewline
G332.4--00.4  & 2000--4000  & CC  & 3.1  & $3_{-1}^{+8}$  & $7_{-4}^{+4}$  & 8\tabularnewline
G353.9--02.0  & ?  &  & unknown  & $1_{-1}^{+2}$  & $3_{-1}^{+1}$  & 3\tabularnewline
G354.8--00.8  & ?  &  & unknown  & $1_{-1}^{+2}$  & $10_{-1}^{+1}$  & 8\tabularnewline
G356.2+04.5  & ?  &  & unknown  & $1_{-1}^{+2}$  & $5_{-2}^{+3}$  & 3\tabularnewline
G359.1--00.5  & $\ge$10000  &  & 5--8.5  & $1_{-1}^{+2}$  & $3_{-1}^{+1}$  & 8\tabularnewline
\hline 
\hline 
\multicolumn{7}{|c}{Single-limbed}\tabularnewline
\hline 
\hline 
G024.7--00.6  & 9500  &  & unknown  & $7_{-3}^{+4}$  & $7_{-3}^{+4}$  & 3\tabularnewline
G166.0+04.3  & ?  &  & 3--6  & $5_{-1}^{+2}$ {[}or $1_{-1}^{+1}${]}  &  & 4\tabularnewline
G182.4+04.3  & 3800--4400  &  & >3  & $6_{-2}^{+2}$ {[}or $1_{-1}^{+2}${]}  &  & 4\tabularnewline
G315.1+02.7  & ?  &  & 1.7  & $7_{-1}^{+2}$  &  & 9\tabularnewline
G327.4+01.0  & ?  &  & unknown  & $1_{-0.5}^{+2}$  & $7_{-3}^{+3}$  & 8\tabularnewline
G338.1+00.4  & ?  &  & unknown  & $2_{-2}^{+1}$  & $3_{-1}^{+1}$  & 8\tabularnewline
G350.0--02.0  & ?  &  & unknown  & $3_{-1}^{+1}$  & $8_{-3}^{+2}$  & 9\tabularnewline
\hline 
\end{tabular}
\tablebib{
(1) Very Large Array via NRAO Science Data Archive, (2) The NRAO VLA
Sky Survey (NVSS, \citealp{Condon:1998kn}), (3) MAGPIS: A Multi-Array
Galactic Plane Imaging Survey \citep{Helfand:2006kw}, (4) Canadian
Galactic Plane Survey, (CGPS, \citealp{Taylor:2003iz}), (5) \citet{Landecker:1999im},
(6) The Westerbork Northern Sky Survey (WENSS, \citealp{Rengelink:1997fo}),
(7) Sino-German $\lambda6$ cm polarization survey \citep{Gao:2010fu},
(8) The Molonglo Observatory Synthesis Telescope (MOST) Supernova
Remnant Catalogue, \citep{Whiteoak:1996tn}, (9) The Parkes-MIT-NRAO
surveys \citep{Condon:1993kf}. Published ages, SN-types, and distances
are taken from SNRcat \citep[\url{http://www.physics.umanitoba.ca/snr/SNRcat/}]{Ferrand:2012cr}
and references therein. These values were retrieved from the website
on 29 February 2016. } 
\end{table}

\onecolumn

\begin{appendix}
\section{\label{sub:Data-models}Data shown in comparison to the models} 
\begin{figure*}
\centering
\includegraphics[width=17cm]{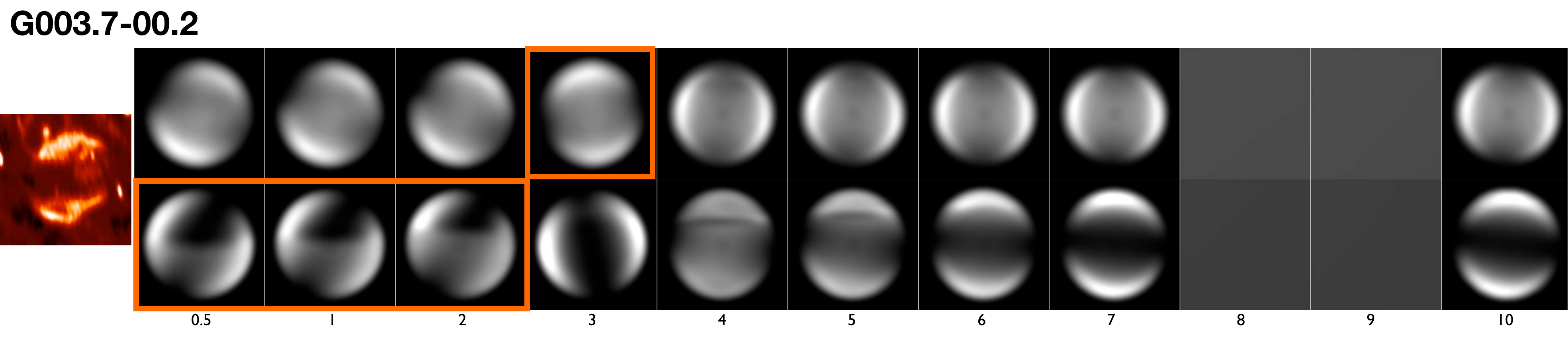}
\includegraphics[width=17cm]{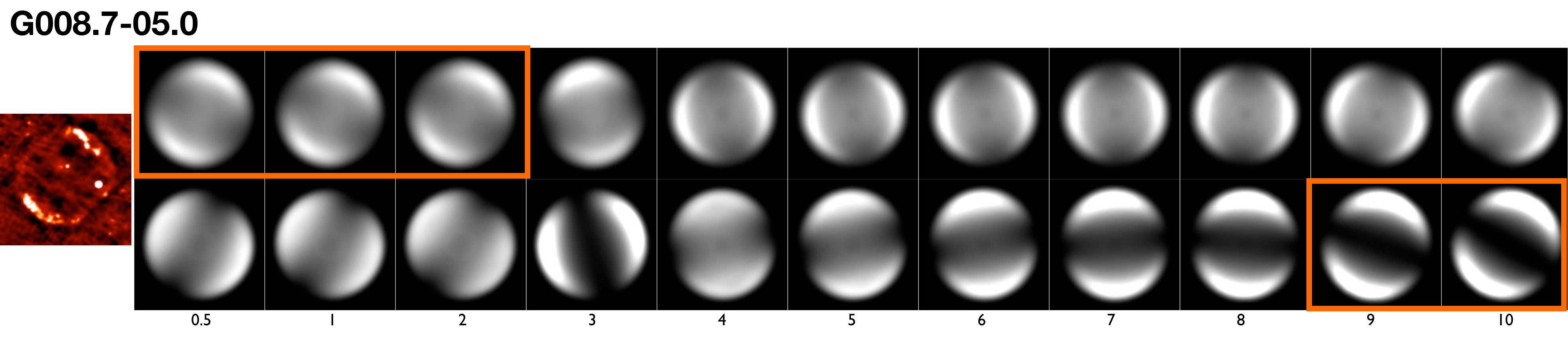}
\includegraphics[width=17cm]{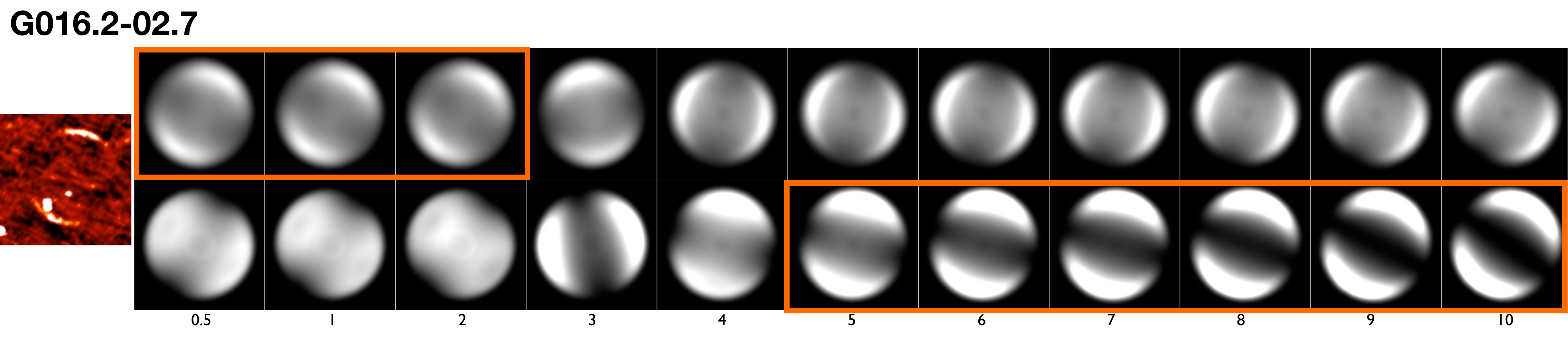}
\includegraphics[width=17cm]{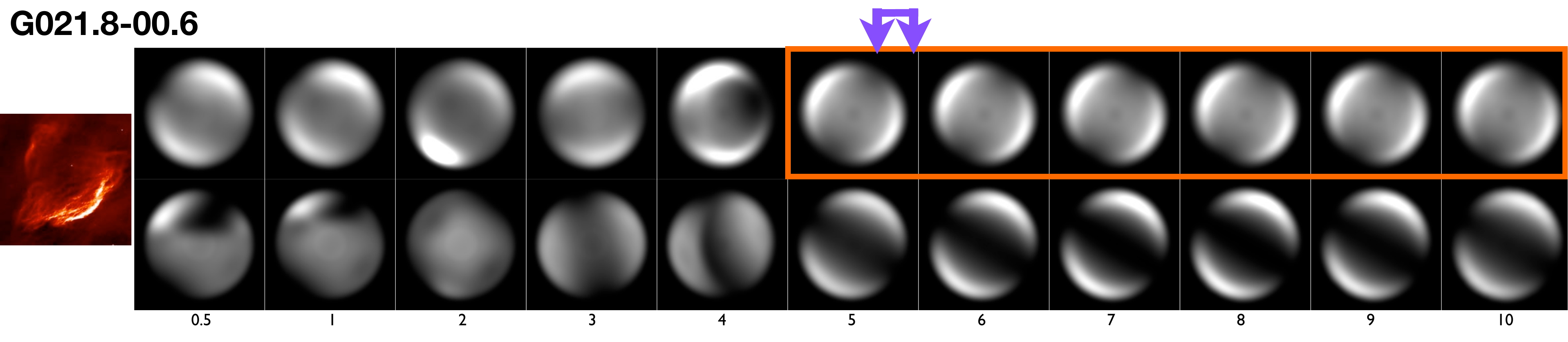}
\includegraphics[width=17cm]{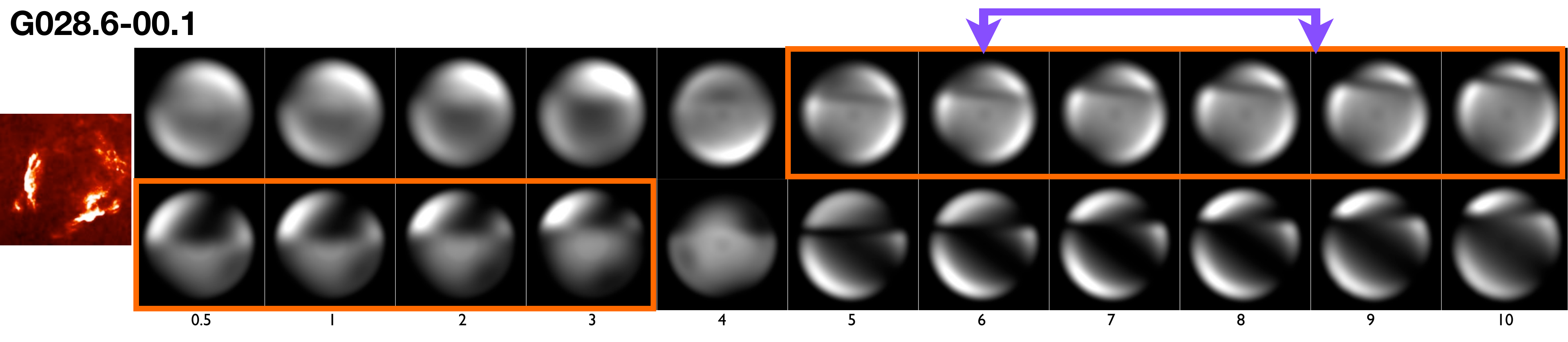}

\protect\caption{\label{fig:data-models-double} Data (left) shown in comparison to models at distances of 0.5, 1, 2, 3, 4, 5, 6, 7, 8, 9, and 10 kpc (left to right) showing the quasi-perpendicular (top) and quasi-parallel (bottom) CRE acceleration case for each SNR in the sample with a double limb. The set of best fitting models, based on the visual appearance of the angle is highlighted with an orange box. In some cases the model is undefined at a location and so the model image will show blank. Where a published value for the distance is available, the range is indicated by an arrow above the models (references for these distances are summarized in Table 1).
}

\end{figure*}

In Figures \ref{fig:data-models-double} and  \ref{fig:data-models-single}, we present models for the cases of quasi-perpendicular and quasi-parallel CRE acceleration and compare these to images for the double-sided and single-sided SNRs respectively. In each case the data are shown on the left (image references are summarized in
Table \ref{tab:data-vs-model}). 

To the right of the image are two strips of models: the models for the quasi-perpendicular CRE
distribution are shown on the top row and the models for the quasi-parallel CRE
distribution are shown below. 
These were made for the position of the particular SNR and at the various
distances as labeled (in kpc). In some cases the Galactic field model
is undefined at a location and so the model image will show blank (see Paper 1).
The set of best fitting models, based on the visual
appearance of the angle is highlighted with an orange box. In almost one-third of the quasi-parallel cases (G021.8--00.6, G036.6+02.6, G046.8--00.3, G054.4--00.3, G166.0+04.3, G182.4+04.3, G302.3+00.7, G315.1+02.7, G317.3--00.2, and G321.9--00.3) none of the models were
judged to have a convincing morphological counterpart and so no model was chosen. Where a published value for the distance is available,
the range is indicated by an arrow above the models.

\begin{figure*}
\centering
\includegraphics[width=17cm]{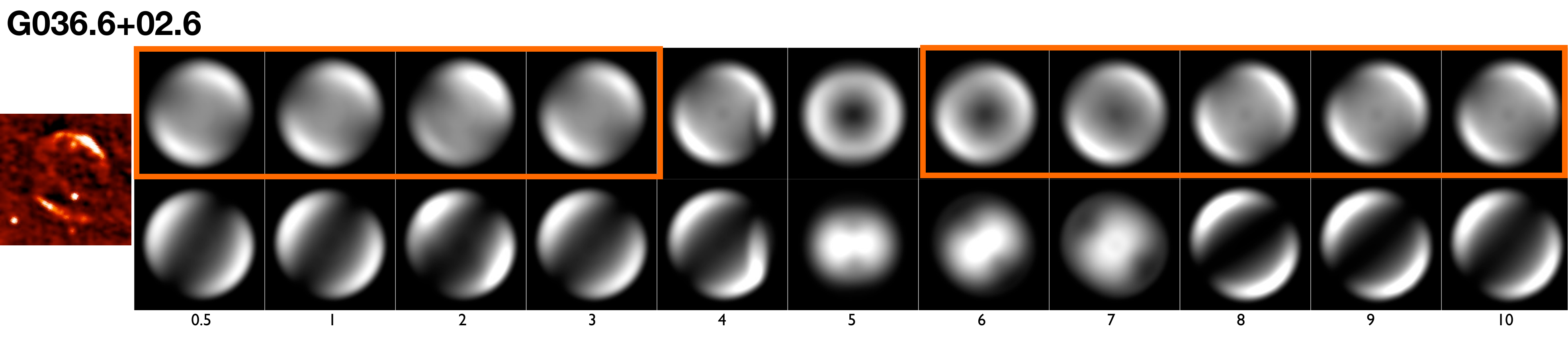}
\includegraphics[width=17cm]{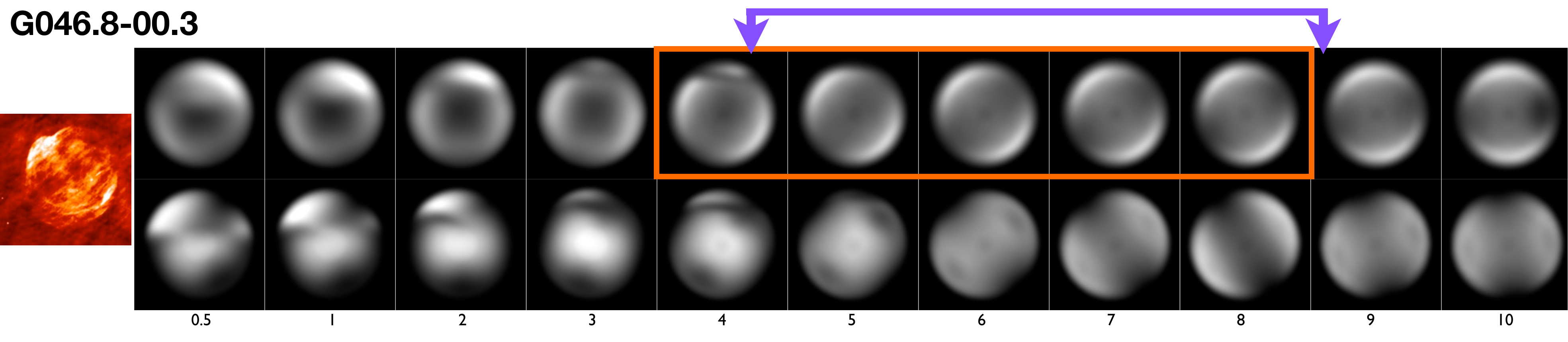}
\includegraphics[width=17cm]{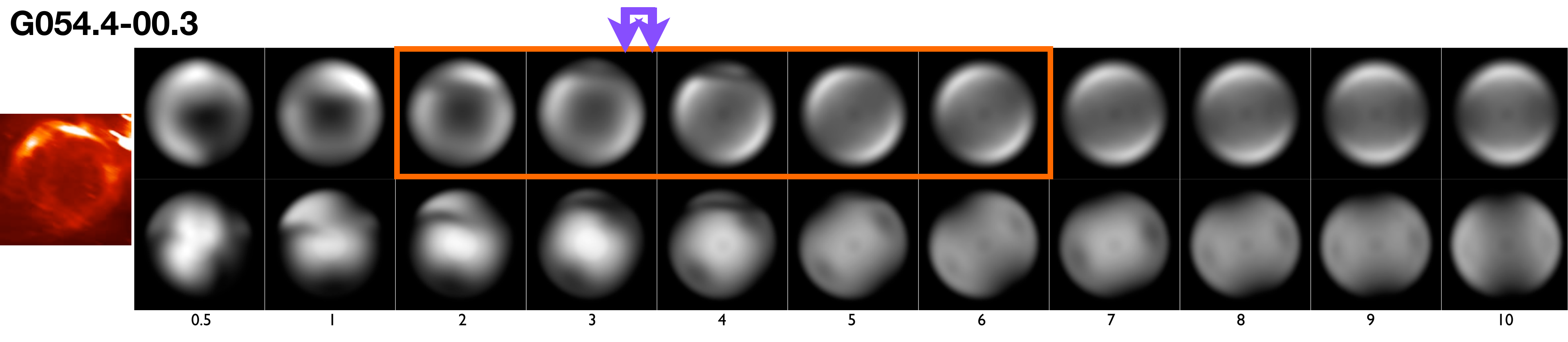}
\includegraphics[width=17cm]{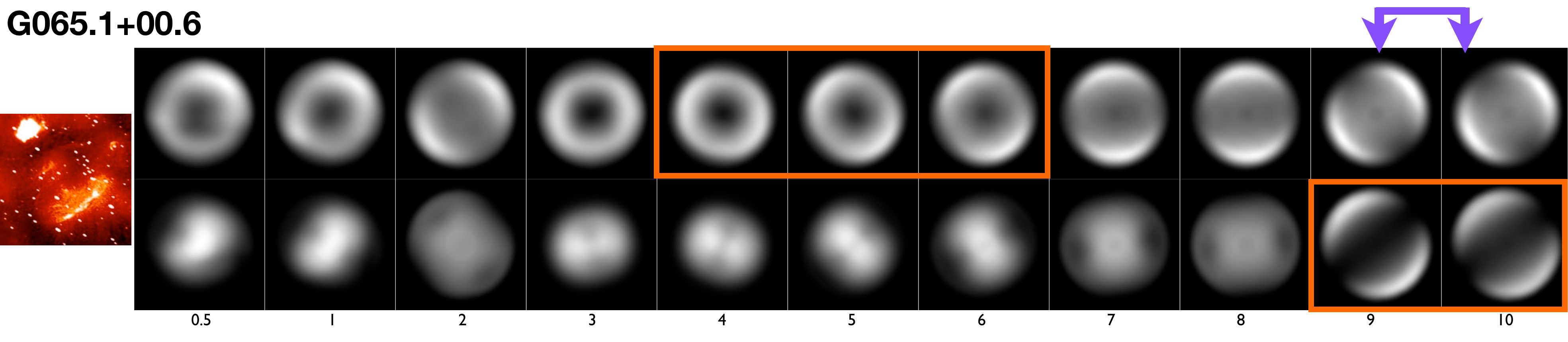}
\includegraphics[width=17cm]{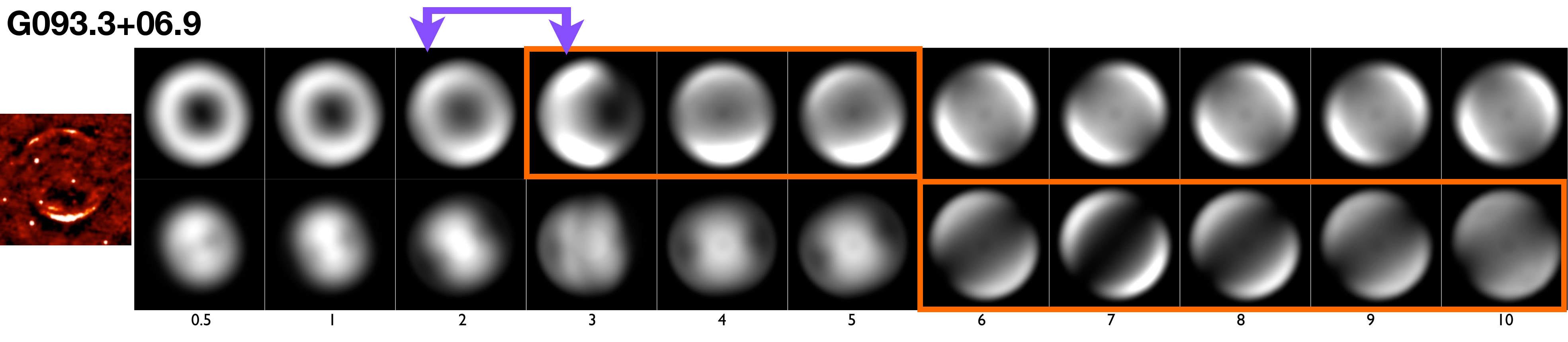}
\includegraphics[width=17cm]{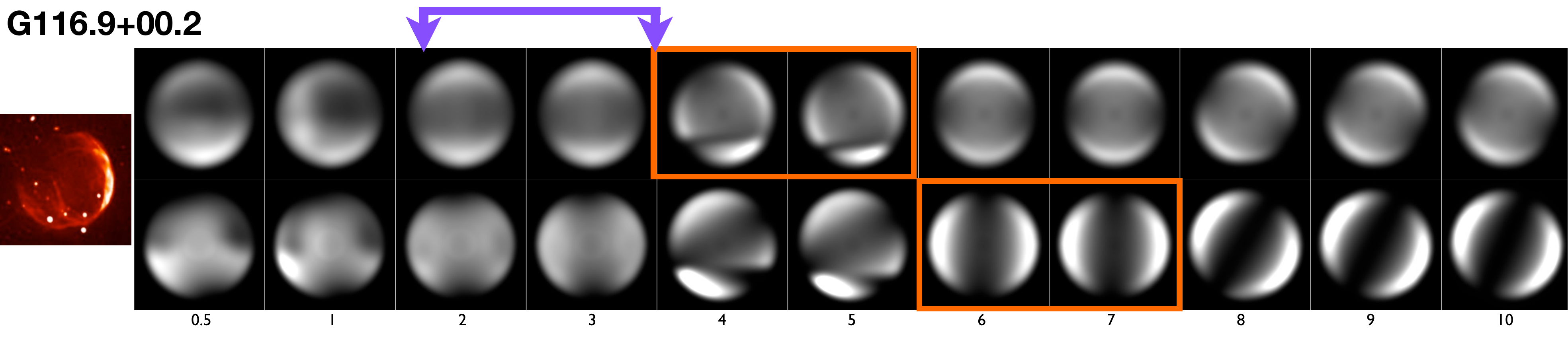}
\\Fig. \ref{fig:data-models-double} continued.
\end{figure*}

\begin{figure*}
\centering
\includegraphics[width=17cm]{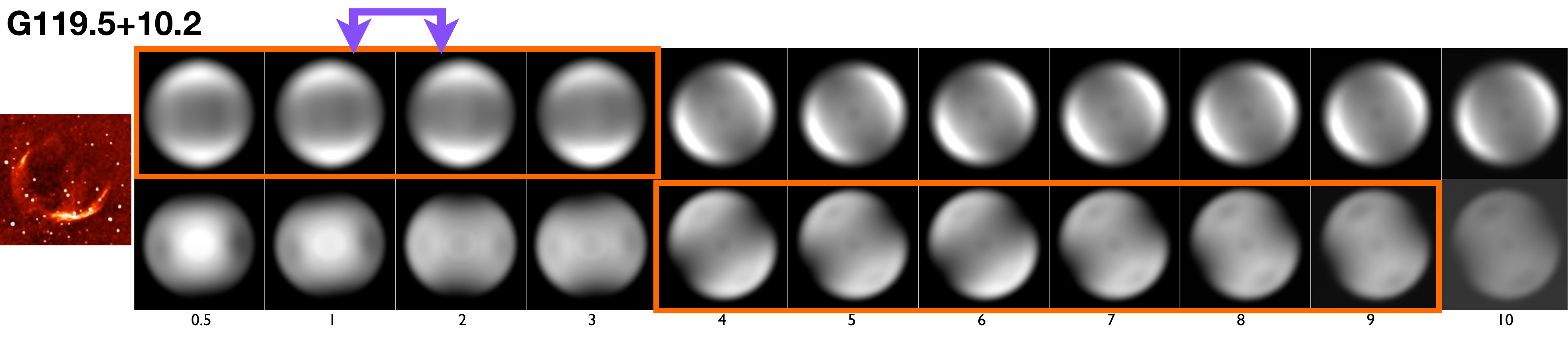}
\includegraphics[width=17cm]{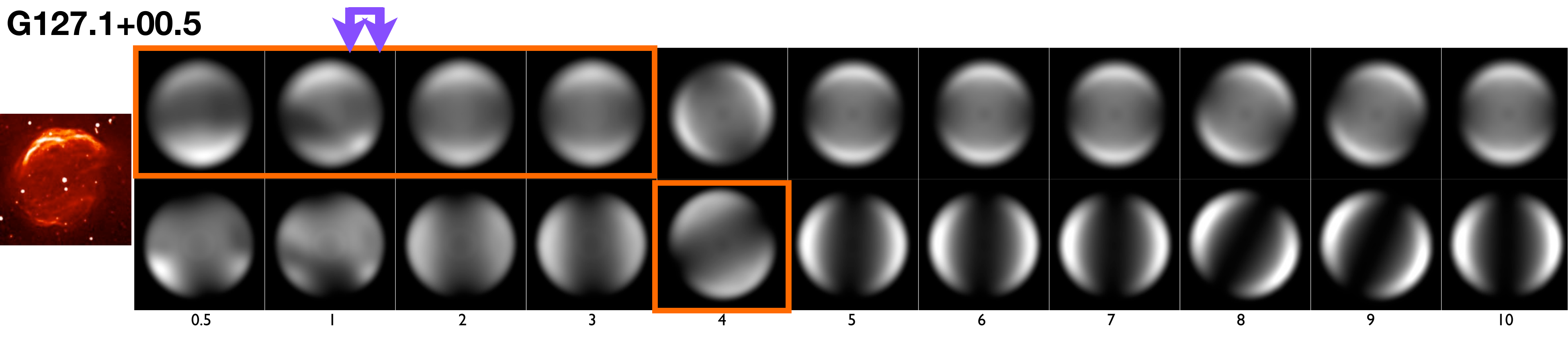}
\includegraphics[width=17cm]{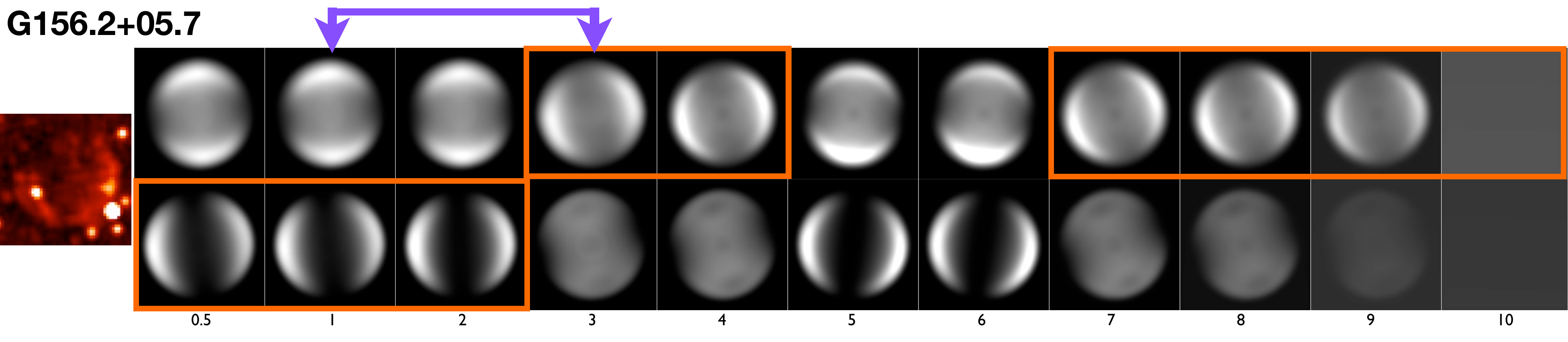}
\includegraphics[width=17cm]{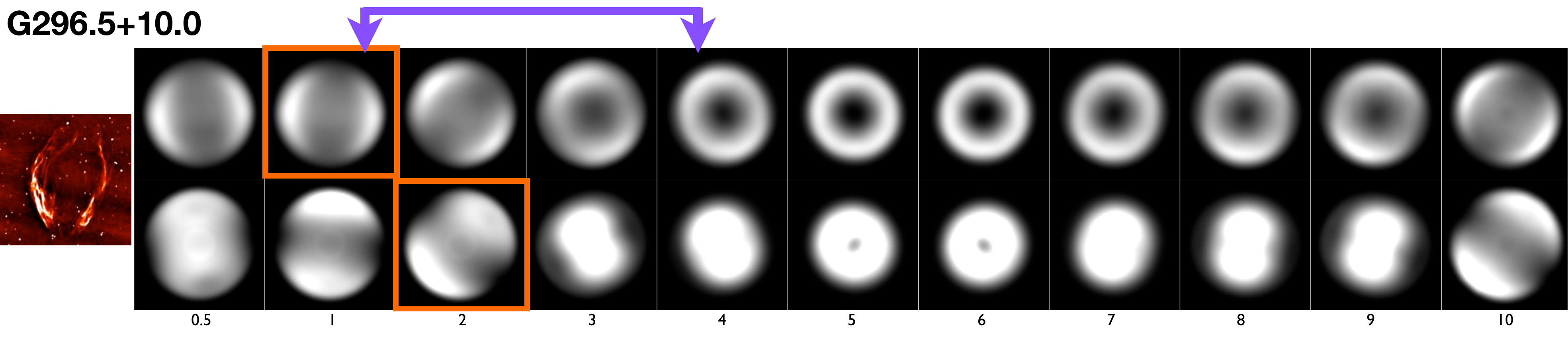}
\includegraphics[width=17cm]{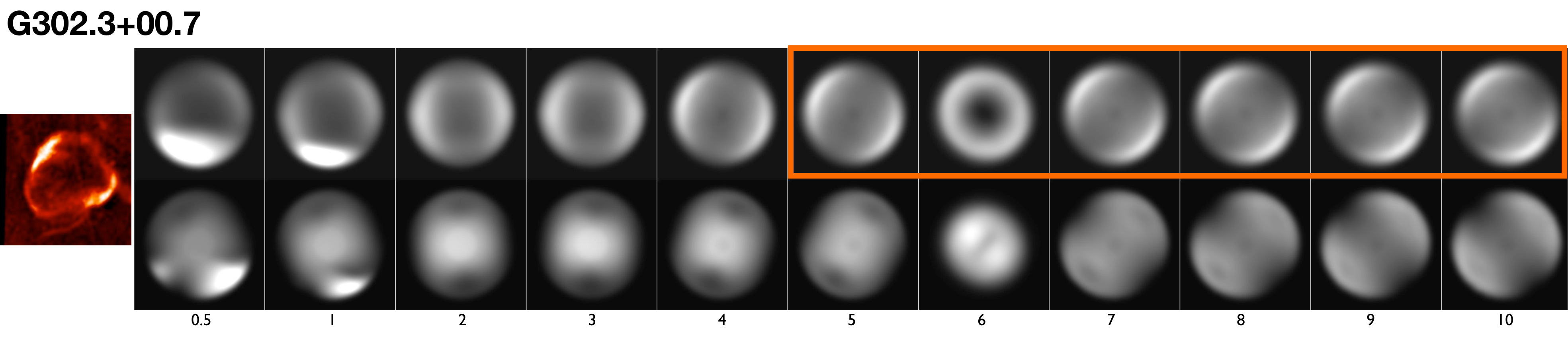}
\includegraphics[width=17cm]{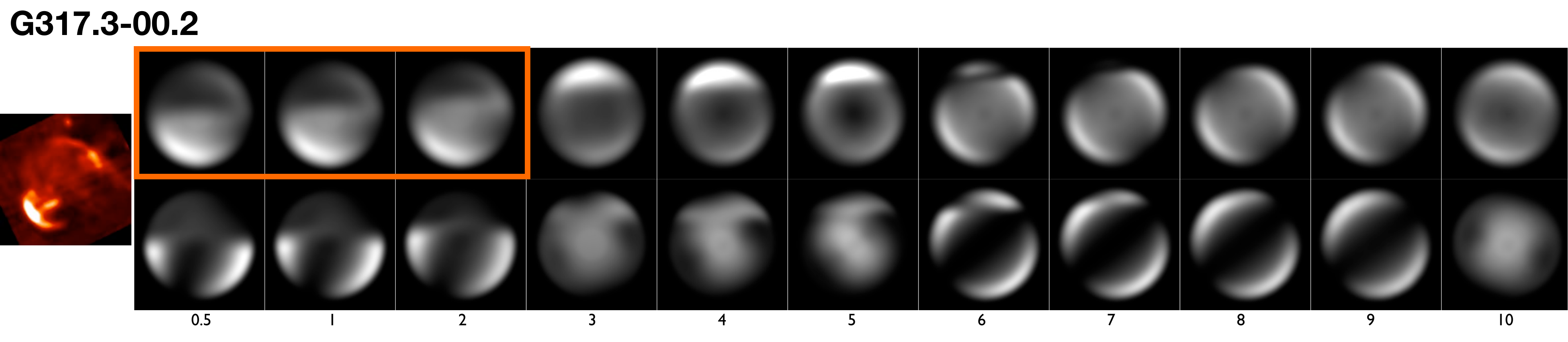}
\\Fig. \ref{fig:data-models-double} continued.
\end{figure*}
\begin{figure*}
\centering
\includegraphics[width=17cm]{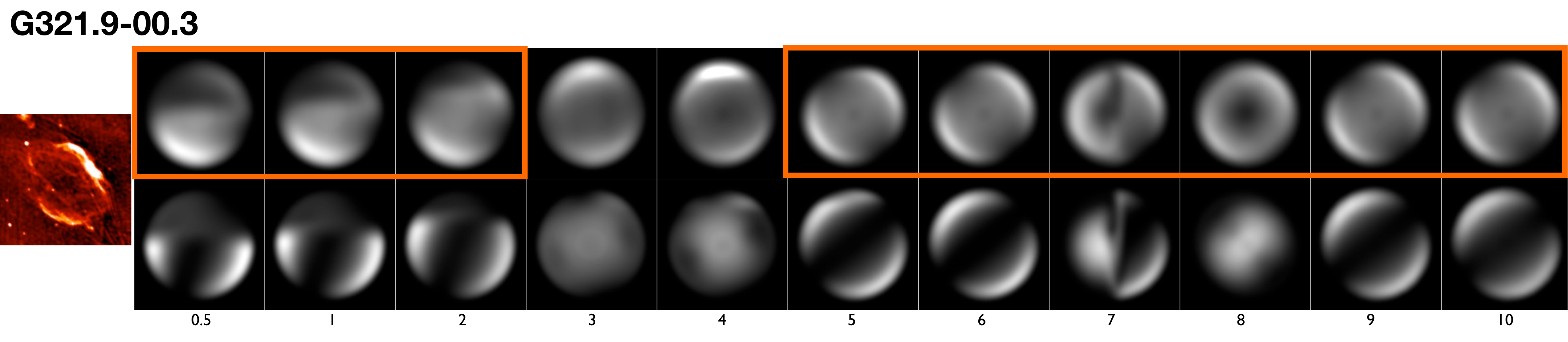}
\includegraphics[width=17cm]{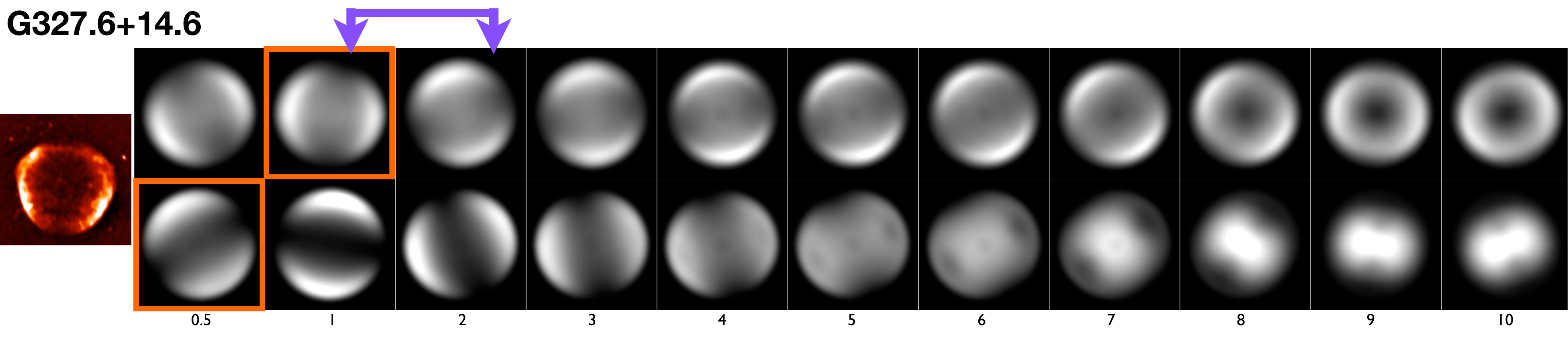}
\includegraphics[width=17cm]{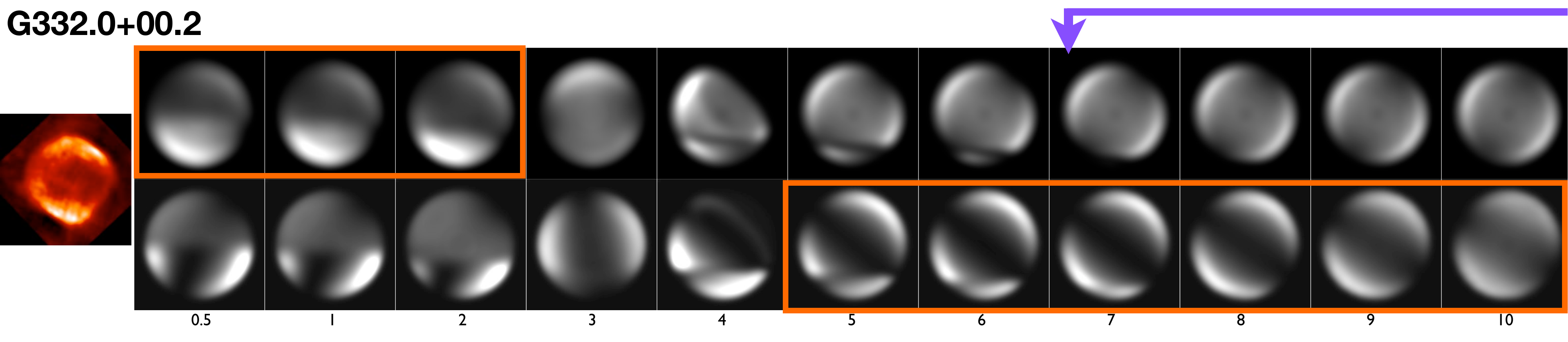}
\includegraphics[width=17cm]{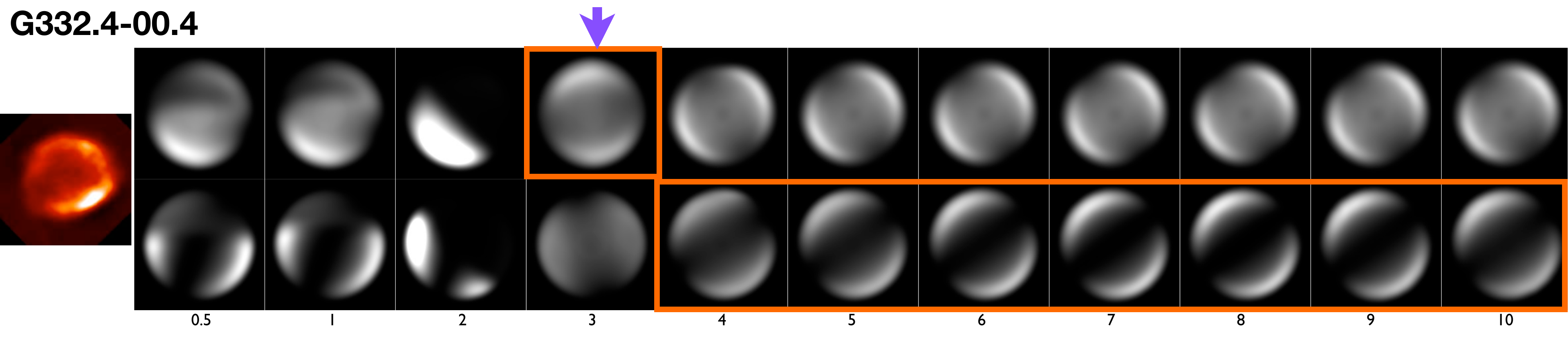}
\includegraphics[width=17cm]{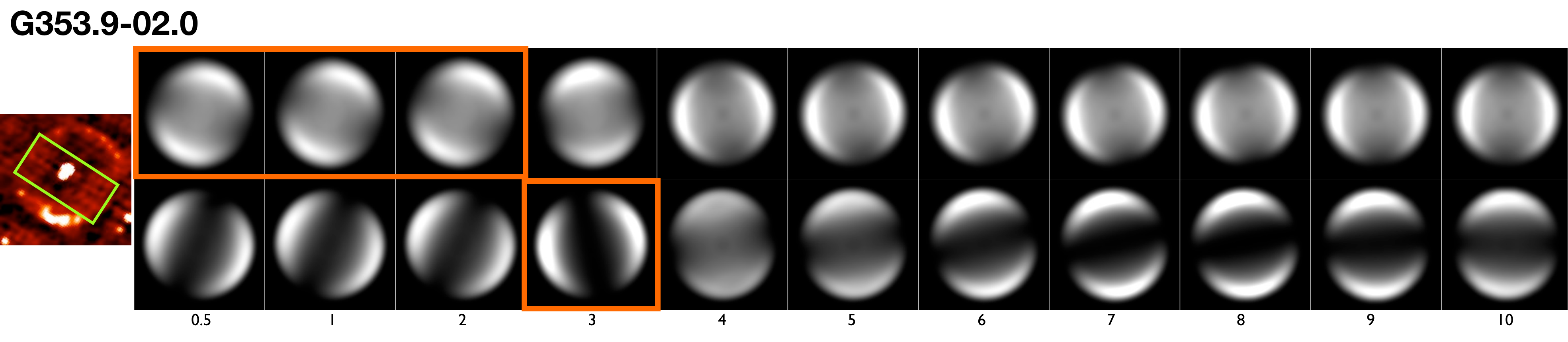}
\includegraphics[width=17cm]{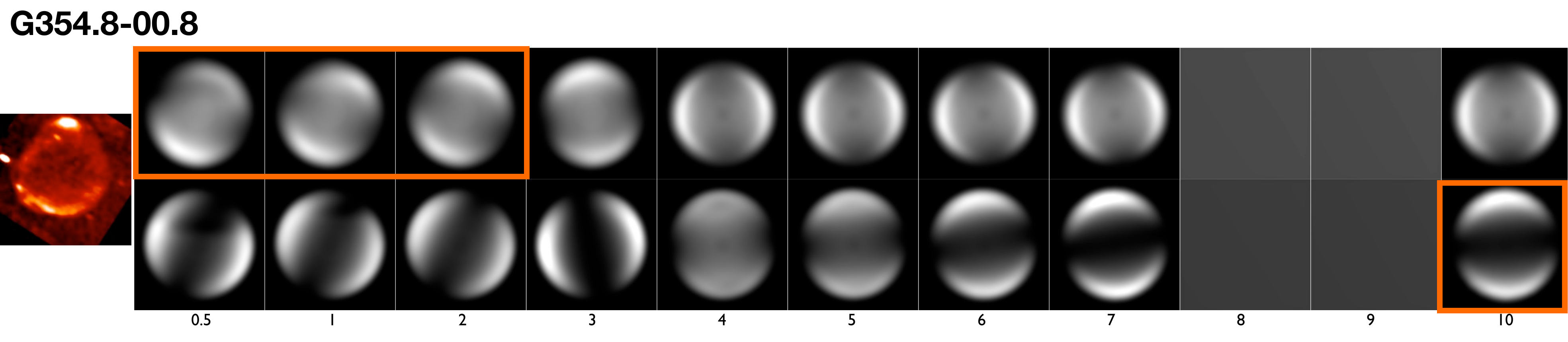}
\\Fig. \ref{fig:data-models-double} continued.
\end{figure*}
\begin{figure*}
\centering
\includegraphics[width=17cm]{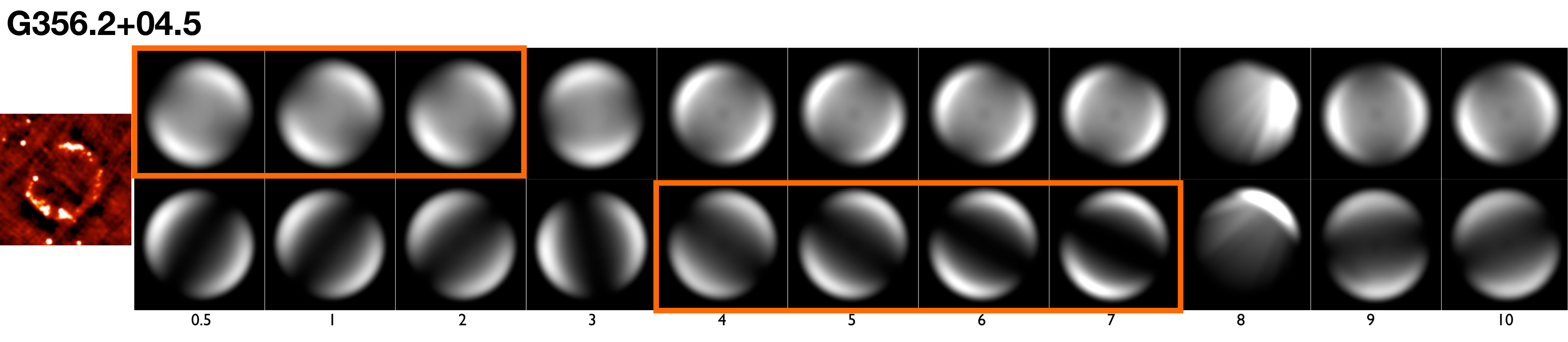}
\includegraphics[width=17cm]{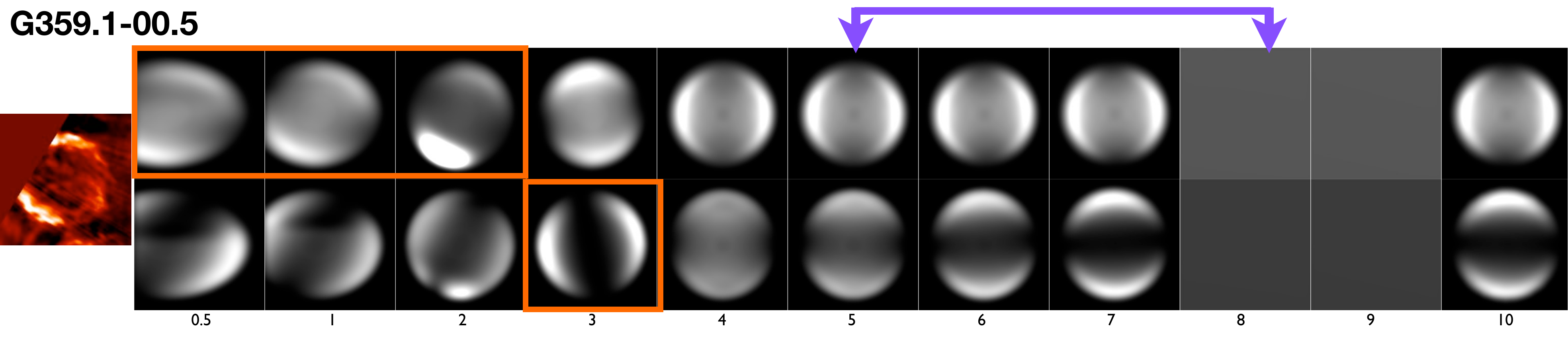}
\\Fig. \ref{fig:data-models-double} continued.
\end{figure*}

\begin{figure*}
\centering
\includegraphics[width=17cm]{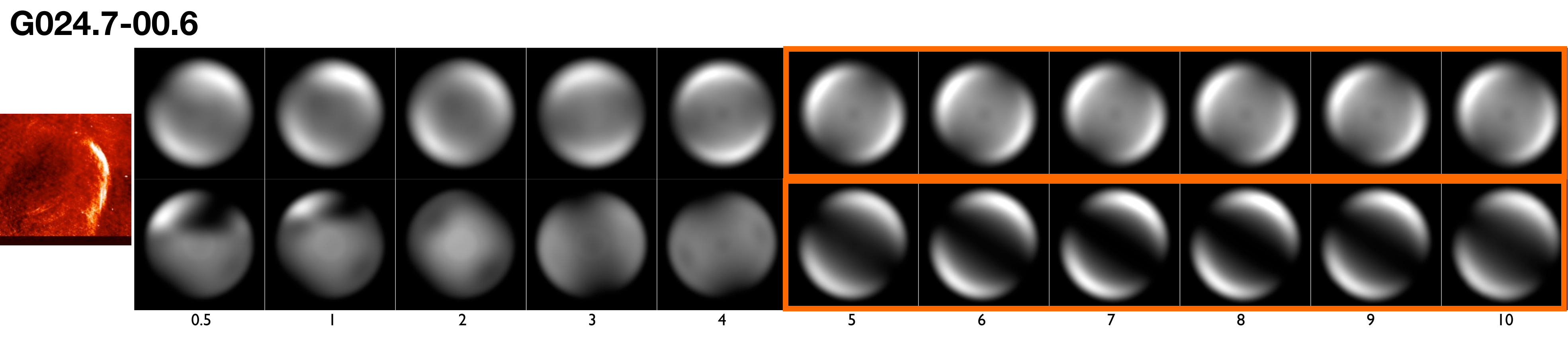}
\includegraphics[width=17cm]{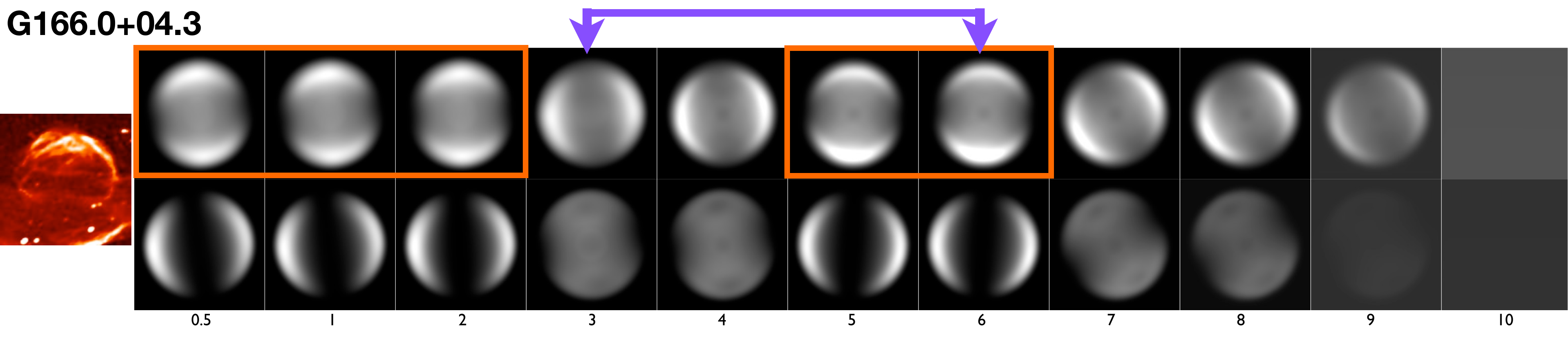}
\includegraphics[width=17cm]{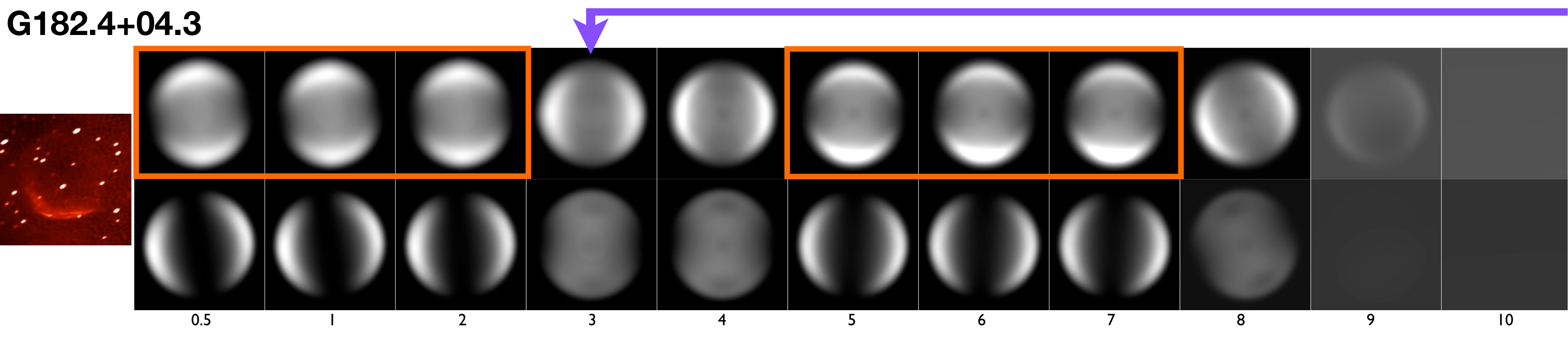}

\protect\caption{\label{fig:data-models-single} Data (left) shown in comparison to models at distances of 0.5, 1, 2, 3, 4, 5, 6, 7, 8, 9, and 10 kpc (left to right) showing the quasi-perpendicular (top) and quasi-parallel (bottom) CRE acceleration case for each SNR in the sample with a single limb. In some cases the model is undefined at a location and so the model image will show blank. Where a published value for the distance is available, the range is indicated by an arrow above the models (references for these distances are summarized in Table \ref{tab:data-vs-model}).
}

\end{figure*}
\begin{figure*}
\centering
\includegraphics[width=17cm]{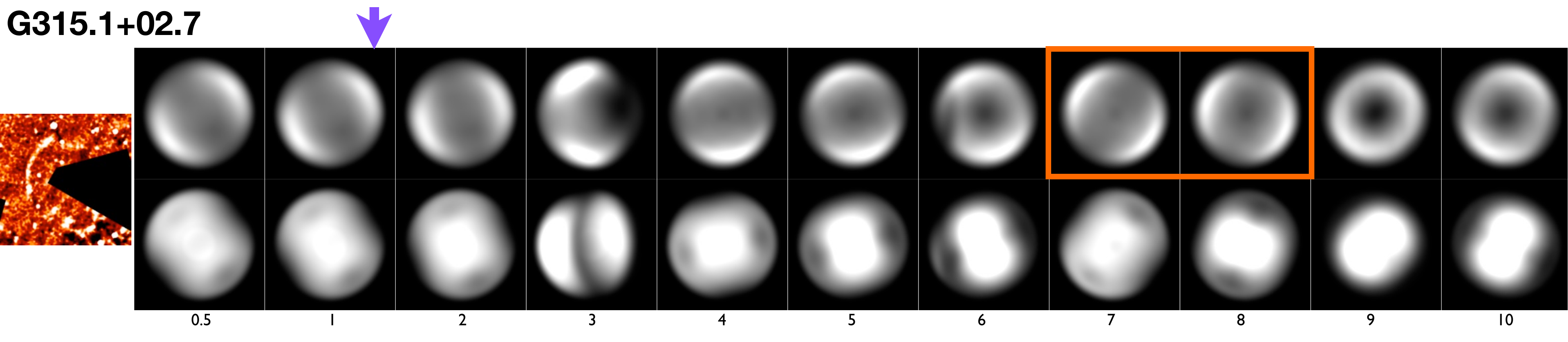}
\includegraphics[width=17cm]{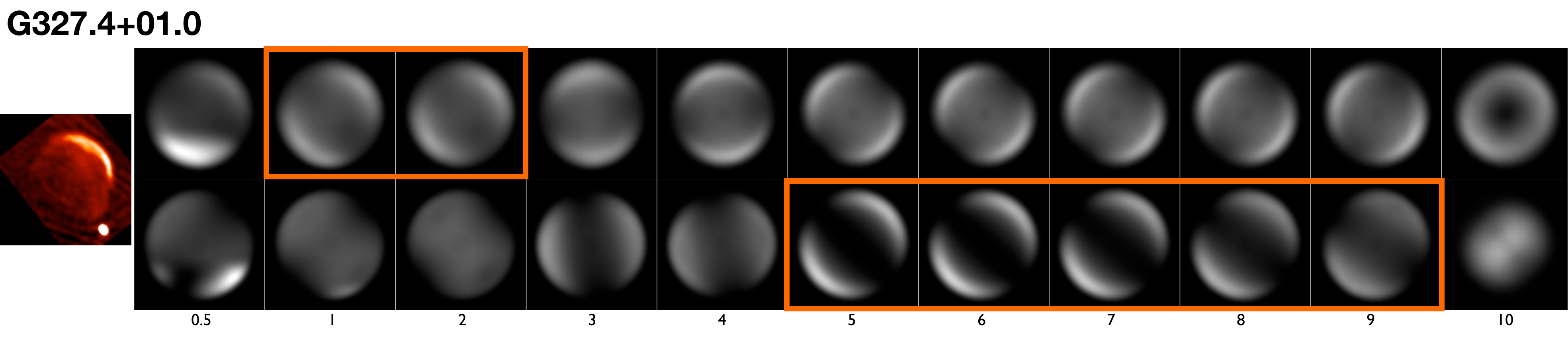}
\includegraphics[width=17cm]{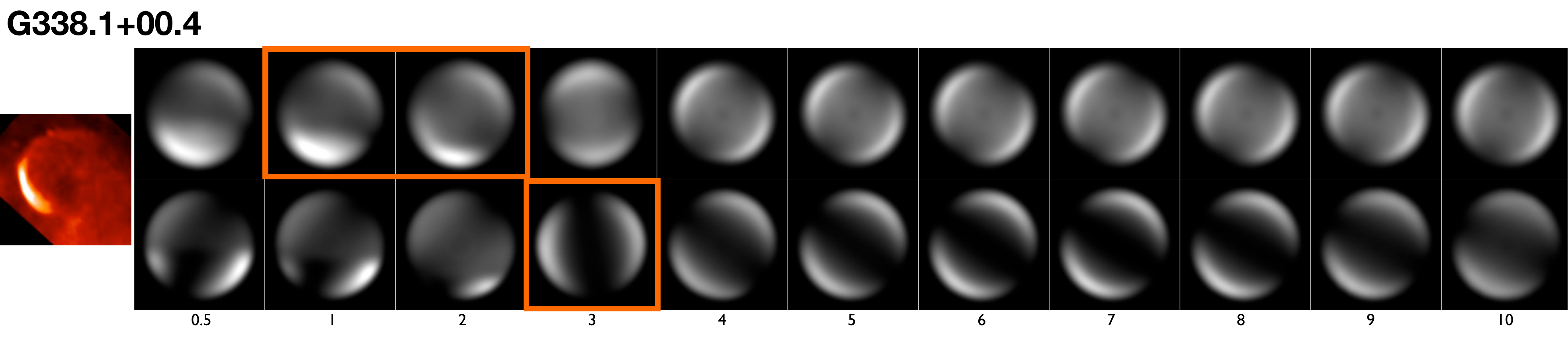}
\includegraphics[width=17cm]{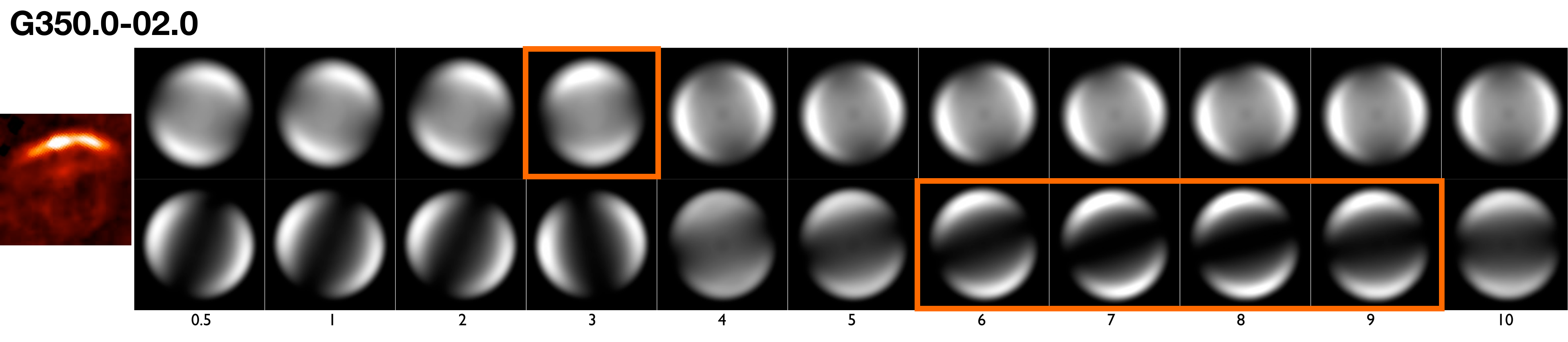}
\\Fig. \ref{fig:data-models-single} continued.
\end{figure*}

\end{appendix}

\end{document}